\documentclass[
twocolumn,prd,
showpacs,
nofootinbib,
amsmath,amssymb,
superscriptaddress]{revtex4-1}

\usepackage{float}
\usepackage{graphicx}% Include figure files
\usepackage{dcolumn}% Align table columns on decimal point
\usepackage{bm}% bold math
\usepackage{hyperref}% add hypertext capabilities
\usepackage{ulem} %xout command
\usepackage{gensymb}
\usepackage{subcaption}
\newcommand{\be}{\begin{equation}}
\newcommand{\ee}{\end{equation}}
\newcommand{\bi}{\begin{itemize}}
\newcommand{\ei}{\end{itemize}}
\newcommand{\bea}{\begin{eqnarray}}
\newcommand{\eea}{\end{eqnarray}}

\usepackage{color}

\usepackage{soul}

\def\newacronym#1#2#3{\gdef#1{#3 (#2)\gdef#1{#2}}}

\def\prl{Physical Review Letters }    % Physical Review Letters
\newacronym{\sdsc}{SDSC}{San Diego Supercomputer Center}
\newacronym{\cgp}{CGP}{Center for Gravitational Physics}
\newacronym{\TACC}{TACC}{Texas Advanced Computing Center}
\newacronym{\nr}{NR}{numerical relativity}
\newacronym{\DA}{DA}{data analysis}
\newacronym{\PCA}{PC}{Principal Component}
\newacronym{\CBC}{CBC}{compact object coalescences}
\newacronym{\grmhd}{GRMHD}{general relativistic magneto-hydrodynamic}
\newacronym{\mhd}{MHD}{magneto-hydrodynamic}
\newacronym{\ornl}{ORNL}{Oak Ridge National Laboratory}
\newacronym{\lisa}{LISA}{the Laser Interferometer Space Antenna}
\newacronym{\lsc}{LSC}{LIGO Scientific Collaboration}
\newacronym{\lvk}{LVK}{LIGO Scientific, Virgo and KAGRA Collaboration}
\newacronym{\sph}{SPH}{smooth particle hydrodynamics}
\newacronym{\tsi}{TSI}{Terascale Supernova Initiative}
\newacronym{\wmap}{WMAP}{the Wilkinson Microwave Anisotropy Probe}
\newacronym{\cmbr}{CMBR}{cosmic microwave background}
\newacronym{\imbbh}{IMBBH}{intermediate mass binary black hole}
\newacronym{\hpc}{HPC}{High-performance Computing}
\newacronym{\bssn}{BSSN}{Baumgarte Shapiro Shibata Nakamura}
\newacronym{\lvc}{LVC}{LIGO Virgo Collaboration}
\newacronym{\EOB}{EOB}{Effective One Body}
\newacronym{\tat}{TAT}{Theoretische Astrophysik T\"ubingen}
\newacronym{\ninja}{NINJA}{Numerical INJection Analysis}
\newacronym{\nrar}{NRAR}{Numerical Relativity Analytical Relativity}
\newacronym{\CTS}{CTS}{Conformal Thin Sandwich}
\newacronym{\gr}{GR}{General Relativity}
\newacronym{\ml}{ML}{machine learning}
\newacronym{\uta}{UT-Austin}{University of Texas at Austin}

\def\cbc#1{compact object coalescence#1 (CBC#1)\gdef\cbc{CBC}}
\def\ahz#1{apparent horizon#1 (AH#1)\gdef\ahz{AH}}
\def\qnm#1{quasi-normal mode#1 (QNM#1)\gdef\qnm{QNM}}
\def\snr#1{signal-to-noise ratio#1 (SNR#1)\gdef\snr{SNR}}
\def\si#1{Senior Investigator#1 (SI#1)\gdef\si{SI}}
\def\emri#1{Extreme Mass-Ratio Inspiral#1 (EMRI#1)\gdef\emri{EMRI}}
\def\imbh#1{intermediate mass black hole#1 (IMBH#1)\gdef\imbh{IMBH}}
\def\imbhb#1{intermediate mass black hole binary#1 (IMBHB#1)\gdef\imbhb{IMBHB}}
\def\smbh#1{supermassive black hole#1 (SMBH#1)\gdef\smbh{SMBH}}
\def\bbh#1{binary black hole#1 (BBH#1)\gdef\bbh{BBH}}
\def\bh#1{black hole#1 (BH#1)\gdef\bh{BH}}
\def\ns#1{neutron star#1 (NS#1)\gdef\ns{NS}}
\def\hmns#1{hypermassive neutron star#1 (HMNS#1)\gdef\hmns{HMNS}}
\def\whd#1{white dwarf#1 (WD#1)\gdef\whd{WD}}
\def\gw#1{Gravitational wave#1 (GW#1)\gdef\gw{GW}}
\def\isco#1{innermost stable circular orbit#1 (ISCO#1)\gdef\isco{ISCO}}
\def\EM#1{electromagnetic#1 (EM#1)\gdef\EM{EM}}
\def\pn#1{post-Newtonian#1 (PN#1)\gdef\pn{PN}}
\def\eos#1{equation of state#1 (EOS#1)\gdef\eos{EOS}}
\def\grb#1{gamma-ray burst#1 (GRB#1)\gdef\grb{GRB}}
\def\tde#1{tidal disruption event#1 (TDE#1)\gdef\tde{TDE}}
\def\pca#1{Principal Component Analysis#1 (PCA#1)\gdef\pca{PCA}}
\def\bhns#1{black hole - neutron star#1 (BHNS#1)\gdef\bhns{BHNS}}
\def\dns#1{double neutron star#1 (DNS#1)\gdef\dns{DNS}}
\def\jbd#1{Jordan-Brans-Dicke#1 (JBD#1)\gdef\jbd{JBD}}
\def\cbc#1{compact binary coalescence#1 (CBC#1)\gdef\cbc{CBC}}
\def\atg#1{alternative theory of gravity#1 (ATG#1)\gdef\atg{ATG}}

\def\rift#1{Rapid parameter estimation via Iterative FiTing#1 (\textit{RIFT}#1)\gdef\rift{\textit{RIFT}}}

\newcommand{\carpet}{\texttt{Carpet}}
\newcommand{\cactus}{\texttt{Cactus}}
\newcommand{\kranc}{\texttt{Kranc}}
\newcommand{\maya}{\texttt{Maya}}

\newcommand{\etk}{\textit{Einstein Toolkit}}

\newcommand{\mayawaves}{\texttt{mayawaves}}

\def\ligo#1{the Laser Interferometer Gravitational-wave Observatory#1 (LIGO#1)\gdef\ligo{LIGO}}
\def\et#1{the Einstein Telescope#1 (ET#1)\gdef\et{ET}}
\def\ce#1{Cosmic Explorer#1 (CE#1)\gdef\ce{CE}}

\def\pycbc#1{PyCBC#1\gdef\pycbc{PyCBC}}

\def\sxs#1{Simulating eXtreme Spacetimes#1 (SXS#1) Collaboration#1\gdef\sxs{SXS}}

\def\com#1{center-of-mass#1 (COM#1)\gdef\com{COM}}

\begin{document}

\title{Second MAYA Catalog of Binary Black Hole Numerical Relativity Waveforms}

\affiliation{Center for Gravitational Physics and Department of Physics, The University of Texas at Austin, Austin, TX 78712}
\affiliation{Department of Physics, University of Illinois Urbana-Champaign, Urbana, IL 61801}

\author{Deborah Ferguson$^{1,2}$}\noaffiliation
\author{Evelyn Allsup$^{1}$}\noaffiliation
\author{Surendra Anne$^{1}$}\noaffiliation
\author{Galina Bouyer$^{1}$}\noaffiliation
\author{Miguel Gracia-Linares$^{1}$}\noaffiliation
\author{Hector Iglesias$^{1}$}\noaffiliation
\author{Aasim Jan$^{1}$}\noaffiliation
\author{Pablo Laguna$^{1}$}\noaffiliation
\author{Jacob Lange$^{1}$}\noaffiliation
\author{Erick Martinez$^{1}$}\noaffiliation
\author{Filippo Meoni$^{1}$}\noaffiliation
\author{Ryan Nowicki$^{1}$}\noaffiliation
\author{Deirdre Shoemaker$^{1}$}\noaffiliation
\author{Blake Steadham$^{1}$}\noaffiliation
\author{Max L. Trostel$^{1}$}\noaffiliation
\author{Bing-Jyun Tsao$^{1}$}\noaffiliation
\author{Finny Valorz$^{1}$}\noaffiliation
%\preprint{LIGO-xxxxx}
%\pacs{04.80.Nn, 04.25.dg, 04.25.D-, 04.30.-w} 

\begin{abstract}
Numerical relativity waveforms are a critical resource in the quest to  deepen our understanding of the dynamics of, and gravitational waves emitted from, merging binary systems.
We present 181 new numerical relativity simulations as the second MAYA catalog of binary black hole waveforms (a sequel to the Georgia Tech waveform catalog). 
Most importantly, these include 55 high mass ratio ($q>=4$), 48 precessing,  and 92 eccentric ($e>0.01$) simulations,  including seven simulations which are both eccentric and precessing.
With these significant additions, this new catalog fills in considerable gaps in existing public numerical relativity waveform catalogs. 
The waveforms presented in this catalog are shown to be convergent and are consistent with current gravitational wave models.  
They are available to the public at \url{https://cgpstorage.ph.utexas.edu/waveforms}.
\end{abstract}

\maketitle

\section{Introduction}\label{sec:introduction}
\gw{} observations have become increasingly frequent, with the \lvk{} reporting 90 detections through the end of their third observing run~\cite{LIGOScientific:2021djp, LIGOScientific:2021usb,LIGOScientific:2020ibl, LIGOScientific:2018mvr}, including 83-85 likely \bbh{s}, and even more observed in the fourth observing run.
The \bbh{} systems observed by the \lvk{} span a large range of parameters, including mass ratios from equal mass to the possible %$q=9$ 
9:1 mass ratio of GW190814~\cite{LIGOScientific:2020zkf}.
While most of the events are consistent with nonspinning \bh{s}, some show evidence of in-plane spins~\cite{LIGOScientific:2021djp}, and some  events even show signs of eccentricity~\cite{Gayathri:2020coq, Romero-Shaw:2021ual}.
As we prepare for next-generation \gw{} detectors such as \lisa{}~\cite{2019arXiv190706482B, 2017arXiv170200786A, Robson_2019}, \et{}~\cite{Punturo:2010zz, 2011einstein}, and \ce{}~\cite{2021arXiv210909882E}, we anticipate a significant increase in the volume of signals as well as a more diverse parameter space. 
Understanding the properties of the systems these signals come from and their underlying populations~\cite{LIGOScientific:2020kqk} and  using the signals to test \gr{}~\cite{LIGOScientific:2021sio} relies upon having accurate predictions of the anticipated signals.

Due to the complex nature of Einstein's theory of \gr{}, analytic solutions don't exist for the full two body problem of merging compact objects.
A number of techniques have been developed to solve for the dynamics and radiation of such systems, and the most appropriate technique depends upon the parameters of the binary system in question.
When the objects are far apart, \pn{} approximations can be used with high accuracy and confidence~\cite{Isoyama:2020lls, Blanchet:2013haa, Schafer:2018kuf, Futamase:2007zz, Blanchet:2008je, Faye:2012we, Faye:2014fra}.
For systems where the masses of the black holes are highly unequal, small mass ratio approximations allow for efficient simulations~\cite{Hinderer:2008dm, Miller:2020bft, Barack:2018yvs}.
Great progress has also been made towards using small mass ratio approximations for the inspiral of near equal mass ratio systems~\cite{vandeMeent:2020xgc, Wardell:2021fyy,Warburton:2021kwk, Albertini:2022rfe, Albertini:2022dmc}.

In the highly nonlinear regime of the merger of comparably massed compact objects, \nr{} is the most accurate approach.
\nr{} computationally solves Einstein's equations by expressing them in a 3+1 formalism and separating the constraint and evolution equations in order to create an initial value problem that can be evolved on a computer~\cite{Baumgarte:1998te, baumgarte_shapiro_2010}.
For the case of \bbh{s} in vacuum, there is no missing physics within the simulations, as long as GR is correct, and the only limitations are computational in nature. 
An \nr{} simulation will solve Einstein's equations for a single binary system described by the masses and spin vectors of the two component \bh{s}.
For \bbh{s} in vacuum,  the results of the simulation can be scaled by the total mass, so we set the total ADM mass of the system to 1, and the masses can then simply be described by the mass ratio, $q=m_1/m_2 \geq 1$.
By creating template banks of \nr{} simulations, \nr{} waveforms have been used directly in \gw{} detection and parameter estimation~\cite{Lange:2017wki, LIGOScientific:2016kms, Gayathri:2020coq, CalderonBustillo:2022cja}.

These simulations are placed discretely throughout the parameter space and are time consuming and computationally expensive.
Therefore, while \nr{} waveforms can be used directly as template banks for \gw{} analysis, for many analyses, analytic or semi-analytic models are created, using information from \nr{} for calibration~\cite{Varma:2019csw, Bohe:2016gbl,Khan:2015jqa, Blackman:2017pcm, Husa:2015iqa, Taracchini:2013rva,Hannam:2013oca, Santamaria:2010yb, Nakano:2011pb, Pratten:2020ceb, Ramos-Buades:2023ehm, Colleoni:2024knd, Thompson:2023ase, Nagar:2018zoe}. 
However, creating reliable models and template banks relies upon \nr{} waveforms being sufficiently accurate and densely covering the complete parameter space~\cite{Ferguson:2020xnm, Purrer:2019jcp}.

There are several \nr{} codes and public waveform catalogs created by the \nr{} community.
The SXS collaboration has a public waveform catalog generated using their SpEC code~\cite{2013PhRvL.111x1104M, 2019CQGra..36s5006B} and have been developing their next-generation code, SpECTRE~\cite{deppe_nils_2021_5083825}.  
Additionally, the newest BAM numerical relativity catalog explores the single spin precessing space~\cite{Hamilton:2023qkv, Bruegmann:2006ulg, Hannam:2009rd}.
There are several waveform catalogs generated with codes derived from an open source software used by many \nr{} groups, the \etk{}~\cite{Jani:2016wkt,Healy:2017psd, Healy:2019jyf, Healy:2022wdn, Zlochower:2005bj, PhysRevD.102.104018, Loffler:2011ay}. 
The waveforms in the catalog presented in this paper are also  generated with a code derived from the \etk{},  called \maya{}~\cite{Jani:2016wkt, Herrmann:2006ks, Vaishnav:2007nm, Healy:2009zm, PhysRevD.88.024040}.

We previously released the Georgia Tech waveform catalog~\cite{Jani:2016wkt} using our \maya{} code including 452 waveforms covering the spinning parameter space up to $q=8$ and including nonspinning waveforms up to $q=15$.
These waveforms have been used to study exceptional \lvk{} events~\cite{Lange:2017wki}, the detectability of intermediate-mass \bh{s}~\cite{Jani:2019ffg, PhysRevD.102.044035, LIGOScientific:2021tfm}, as well as post-merger chirps~\cite{CalderonBustillo:2019wwe}, to highlight a few recent studies.
They have also been used to study the different stages of the merger, including finding a relationship between the frequency at merger and the spin of the remnant \bh{}~\cite{Ferguson:2019slp, Healy:2014eua}.
Using the Georgia Tech catalog, it was also found that the \bh{} recoil can be measured from \gw{} observations~\cite{PhysRevLett.121.191102}, a method which has been applied to real events~\cite{CalderonBustillo:2022ldv}.
This catalog has since been rebranded to the MAYA Catalog.%, which the catalog presented in this paper expands upon.

This paper introduces the second MAYA Catalog, expanding beyond the parameter space coverage of the first catalog and introducing a new format. 
This format can be read using the \mayawaves{} python library~\cite{Ferguson:2023mks, Ferguson_mayawaves_2023},  which is described further in Section ~\ref{sec:mayawaves}.  
The new catalog includes a suite of eccentric simulations as well as improved coverage of the high mass ratio parameter space and the precessing parameter space.
In Section ~\ref{sec:codebase}, we describe the \maya{} code used to create this catalog as well as the \mayawaves{} python library.
We then dive into the description of the new catalog in Section ~\ref{sec:catalog_description}.
Section ~\ref{sec:model_comparisons} compares our simulations to other \nr{} waveforms and waveform models, including an eccentric model, and Section ~\ref{sec:final_state} compares the models for the remnant \bh{} parameters to the results obtained from our catalog. 
Finally, in Section ~\ref{sec:com_correction}, we include a discussion of the impact of \com{} drift corrections.

\section{Codebase}\label{sec:codebase}

In this section, we'll describe the code used to perform these \nr{} simulations (\maya{}) as well as the python library used to interact with and analyze the simulations (\mayawaves{}).

\subsection{\maya{} Code}\label{sec:maya_code}

This catalog was constructed from simulations performed using the \maya{} code, a fork of the \etk{} using additional and modified thorns. 
The \etk{} is a finite-differencing code based upon the \cactus{} infrastructure using \carpet{} for box-in-box mesh refinement~\cite{Schnetter:2003rb}.
\maya{} uses the BSSN formulation~\cite{Baumgarte:1998te} to separate Einstein's equations into constraint equations to solve for the initial data and evolution equations to evolve the spacetime.
The initial data consists of the extrinsic curvature and the spatial metric and is constructed using the TwoPunctures method~\cite{Ansorg:2004ds}.
The extrinsic curvature takes the Bowen-York form~\cite{PhysRevD.21.2047}, and the initial spacial metric is conformally flat. 
To set the initial momenta for quasi-circular simulations, we use the \pn{} equations described in ~\cite{Healy:2017zqj}.
We use the moving puncture gauge condition~\cite{Campanelli:2005dd,Baker:2005vv} and use the \kranc{} scripts to generate the thorns for the spacetime evolution~\cite{Husa:2004ip}.

The Weyl scalar ($\Psi_4$) is computed throughout the computational domain (x,y,z) and then interpolated onto several concentric spheres of various radii.
The waves  are then decomposed and stored using the spin-weighted spherical harmonics ($Y_{\ell,m}$) as a basis:
\begin{equation}
RM\Psi_4(t;\Theta, \Phi) = \sum_{\ell, m}\Psi_{4; \ell, m}(t) {}_{-2}Y_{\ell, m}(\Theta, \Phi) \, ,
\end{equation}
where R is the extraction radius, M is the total mass of the binary, and $\Theta$ and $\Phi$ specify the location on the sphere centered at the \com{} of the binary.
We provide the \gw{} data at several extraction radii. 
We also provide the strain, $h(t)$, which is related to $\Psi_4$ by the following:
\begin{equation}
\Psi_4(t) = \ddot{h}(t)_{+} - i \ddot{h}(t)_{\times} \, ,
\end{equation}
where $\ddot{h}(t)$ refers to the second time derivative.
We use fixed-frequency integration to compute $h(t)$ from $\Psi_4$~\cite{Reisswig:2010di}.
The waveforms can also be extrapolated to infinity using the \mayawaves{} library described below. 

For most of the simulations presented in this catalog, the grids are constructed in a consistent way.
There are three sets of nested grids, one centered at the origin, one centered on the larger black hole, and one centered on the smaller black hole.
For the grids centered on the \bh{s}, the finest grids are set to be just larger than the radius of the \bh{s}, with 48 points across the radius.
Larger grids are then constructed, with radii and grid spacing increasing by powers of 2, with the largest radius being at least 400M.
The largest three levels are centered on the origin.
Since this catalog includes some older miscellaneous simulations, some of the earlier simulations do not follow this exact pattern.
The parameter files for all simulations are included in the simulation data, and their grid structure can be read using the \mayawaves{} library described below.

While our previous catalog paper showed convergence of the \maya{} code, we perform additional tests to reestablish the claims here.
Refer to the Appendix for a detailed analysis of the convergence and estimated errors for this catalog.
In summary, we find the convergence order to range from 2.12 to 2.94 and find that our most complicated systems are unlikely to cause significant parameter estimation bias up to a \snr{} of $\rho = 63.8$, while our more standard systems are unlikely to cause significant parameter bias up to $\rho = 543$.
Some of the older simulations included in this catalog were performed with coarser grids (down to 24 points across the radius of the finest grid).
These simulations will be clearly pointed out in the catalog description provided in Sec.~\ref{sec:catalog_description}.
For such simulations, parameter bias could occur above $\rho = 14.7$ or lower for the highest mass ratio cases ($q={7, 15}$).
In these cases, the waveforms should be used for qualitative studies rather than precision analyses.

\subsection{\mayawaves{} Python Library}\label{sec:mayawaves}
The simulations in this catalog are provided in two formats. 
The more comprehensive format consists of an h5 file for each simulation that contains information relating to the compact objects as well as the radiation they emit. 
This h5 file should be read using the \mayawaves{} python library~\cite{Ferguson:2023mks, Ferguson_mayawaves_2023}.
This library provides easy and efficient use of the MAYA catalog and can also be used to analyze any \etk{} simulation.
Storing the simulations in this format enables us to provide more raw data pertaining to radiated quantities as well as the evolution of the compact objects themselves.
The user need only download the desired waveform from our catalog, import \mayawaves{} in their python file, and create a Coalescence object by passing in the simulation's path to the constructor. 
They can then access all information regarding the simulation through the Coalescence object.
The \mayawaves{} library can be installed via pip or from source at https://github.com/MayaWaves/mayawaves.
Documentation and tutorials can be found at https://mayawaves.github.io/mayawaves. 
Refer to ~\cite{Ferguson:2023mks} for more information on this library.

\section{Catalog Description}\label{sec:catalog_description}

\begin{figure*}
  \centering
  \includegraphics[width = \textwidth]{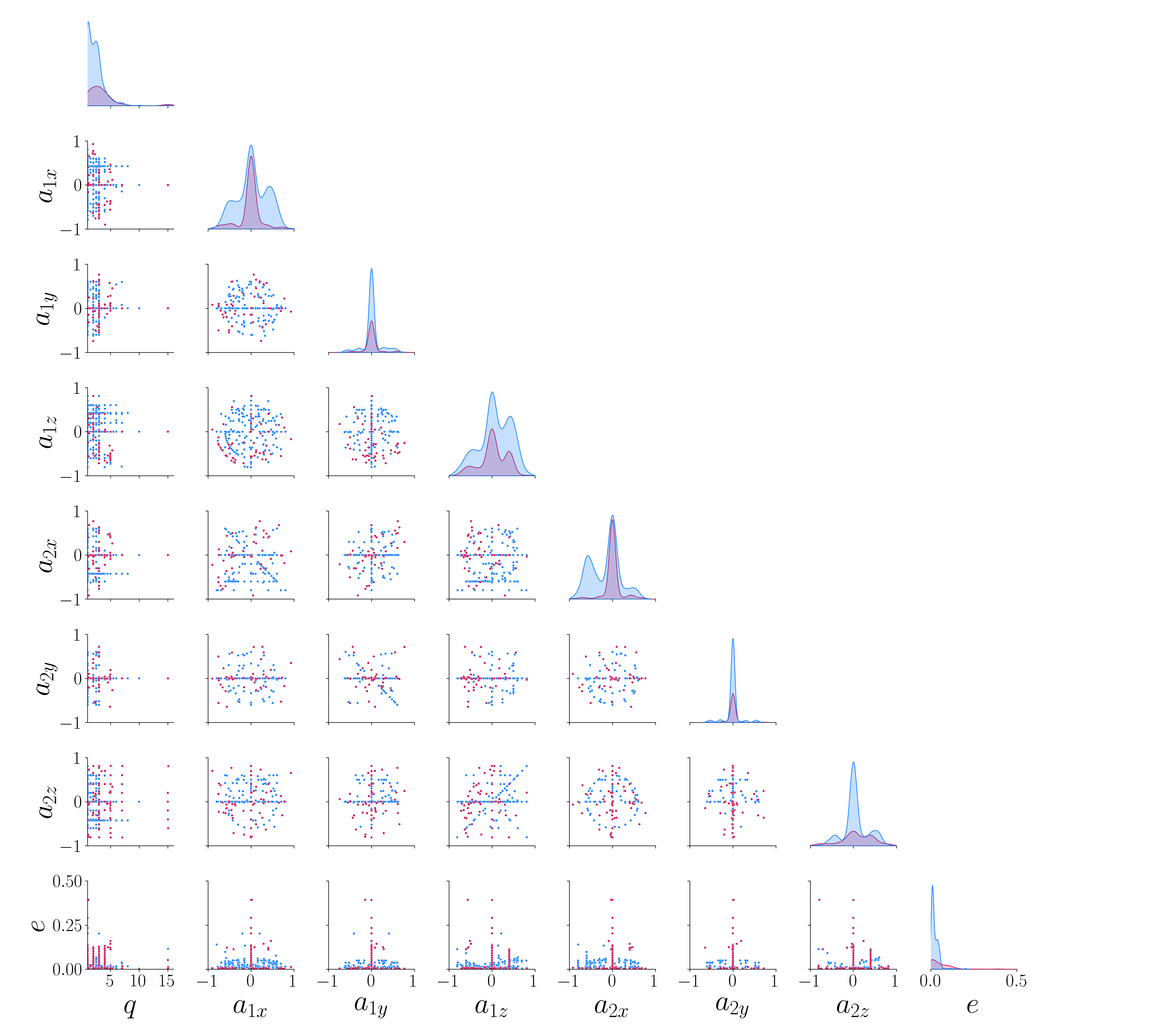}
  \caption{Coverage of the parameter space for the original catalog (blue) and the new catalog (pink). }
  \label{fig:parameter_space}
\end{figure*}

\begin{figure*}
  \centering
  \begin{subfigure}[t]{.48\textwidth}
    \centering
   \includegraphics[width=.95\linewidth]{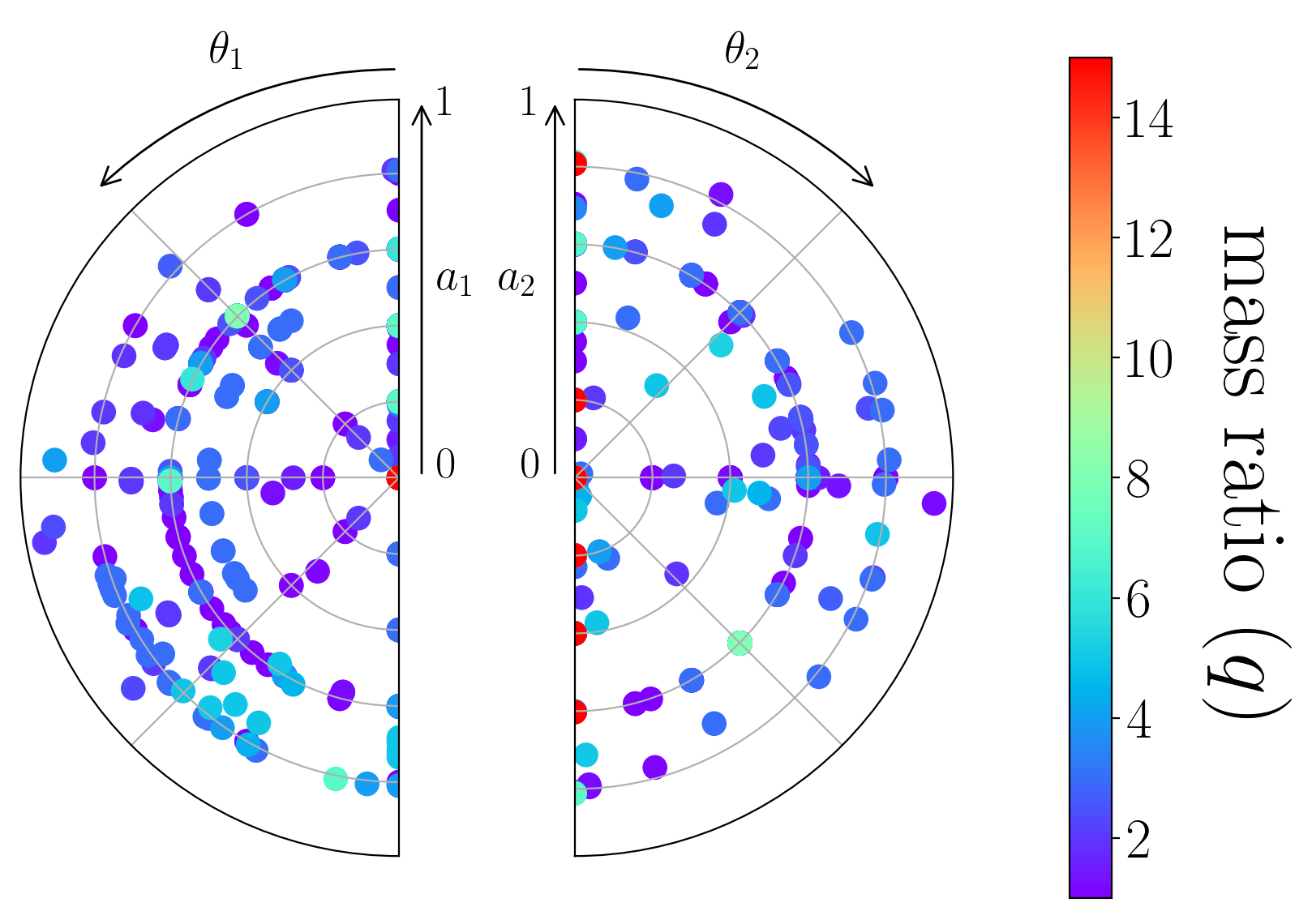}
    \caption{}
    \label{fig:spin_circle_plot}
\end{subfigure}
  \begin{subfigure}[t]{.48\textwidth}
    \centering
   \includegraphics[width=.95\linewidth]{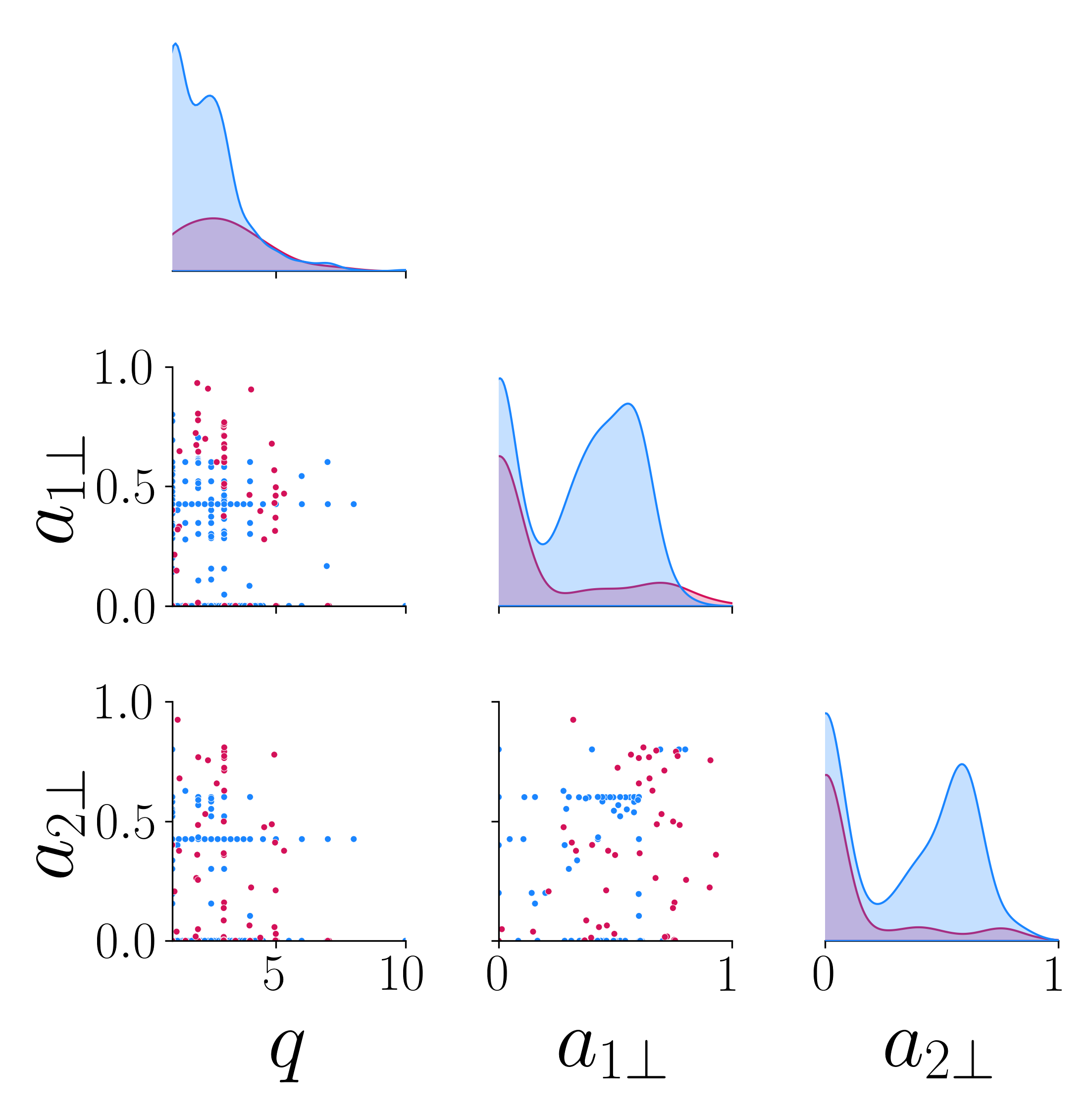}
    \caption{}
    \label{fig:in_plane_corner_plot}
\end{subfigure}
  \caption{Coverage of the spin parameter space.  a) The left semicircle denotes the spin for the primary black hole, with radius showing dimensionless spin magnitude ($a_1$) and the angle from the vertical showing the angle between the spin and the orbital angular momentum ($\theta_1$). The same applies for the secondary black hole ($a_2$ and $\theta_2$), shown on the right semicircle. The color denotes the mass ratio.  b) Corner plot showing the in-plane spins vs the mass ratio for the original catalog (blue) and the new catalog (pink). }
  \label{fig:spin_coverage}
\end{figure*}

\begin{figure}
  \centering
  \includegraphics[width = 0.5 \textwidth]{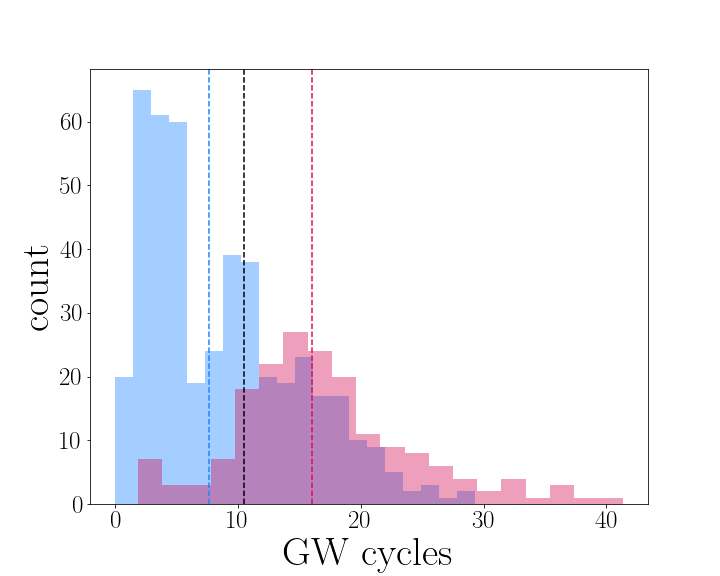}
  \caption{Distribution of the number of inspiral \gw{} cycles for the simulations in the original catalog (blue) and the new catalog (pink).  The vertical lines denote the medians with the black dotted line being the median for the whole catalog.}
  \label{fig:cycles}
\end{figure}

The MAYA catalog of \nr{} waveforms is accessible at \url{https://cgpstorage.ph.utexas.edu/waveforms}.
There, you can access a table with the metadata describing each simulation in the catalog.
From that table, you can download each of the simulations.
They are stored in an h5 format that is designed to be read using the \mayawaves{} python library.
Any simulations that contain all the necessary data are also provided in the format detailed in ~\cite{Schmidt:2017btt} for use with \pycbc{}~\cite{lalsuite, alex_nitz_2020_3630601}.
The waveforms in this format are provided at finite extraction radii.

This catalog introduces 181 new simulations that expand our coverage of the \bbh{} parameter space, for a total of 635 simulations.
The catalog website has also  been updated to include the original catalog~\cite{Jani:2016wkt} in the new \mayawaves{} compatible format.
The initial parameters for all the new catalog simulations are described in Table I at the end of the Appendix.
These parameters are reported at the beginning of the simulation with the initial \bh{s} separated along the x-axis (larger \bh{} at positive x) and the initial orbital angular momentum along the z-axis.
While junk radiation can effect the spins and masses slightly, it has been shown not to be significant~\cite{Higginbotham:2019wbx}.
Therefore, in these plots and tables, we choose the initial pre-junk parameters to maintain the same frame of reference for all simulations.
The only exceptions are that the eccentricity and frequency are provided after the junk radiation as that makes them more accurate.
However, the values will have minimally changed over the 75 M delay it takes junk radiation to settle to $\lesssim 10\%$ of the amplitude of the waveform.
Refer to Sec.~\ref{sec:eccentric} for more information on the computation of the eccentricity.
The values of mass and spin are provided in the h5 files as timeseries so they can be obtained at whatever time/frequency the user chooses.

Figure ~\ref{fig:parameter_space} shows the coverage of the previous catalog and this catalog,  highlighting the many more eccentric simulations, precessing simulations, and high mass ratio simulations introduced by this catalog.
Figure ~\ref{fig:spin_coverage} shows the spin coverage of the combined catalogs.
Figure ~\ref{fig:cycles} shows the distribution of the number of inspiral \gw{} cycles for the simulations in the original and new catalogs, measured from after junk radiation until the peak of the sum of the squares of the $\ell=2$ modes.
Simulations in the new catalog have a median of $\sim 16$ \gw{} cycles and the overall catalog has a median of $\sim 10.5$ \gw{} cycles.

This catalog is an amalgamation of several suites of simulations performed for various studies.
This includes a suite of 80 non-precessing, eccentric simulations with mass ratios up to $q=4$.
It includes 38 simulations for a study of secondary spin for mass ratios up to $q=15$.
We also include nine simulations performed for \lvk{} followup studies.
Finally we have 31 simulations placed to optimally fill in gaps in our parameter space using the neural network and method described in ~\cite{Ferguson:2022qkz}. 
Each of these suites is described in more detail below.
The remaining 23 simulations are an assortment of simulations performed over the last few years.

\subsection{Eccentric}\label{sec:eccentric}

As orbits evolve, eccentricity is quickly radiated away~\cite{PhysRev.136.B1224}; therefore, for most scenarios, the eccentricity is expected to be minimal by the time a stellar mass \bbh{} system reaches current ground-based GW detectors' frequency bands.
As a consequence, most \nr{} simulations and \gw{} models have focused on quasicircular binary systems. 
However, there are astrophysical situations that can lead to non-zero eccentricity in binaries observed by ground-based detectors~\cite{Romero-Shaw:2021ual, PhysRevD.97.103014, PhysRevLett.120.151101, Morscher_2015, Zevin_2019}.  
Some future \gw{} detectors such as \lisa{}, will observe many binaries earlier in their inspiral and are expected to see some eccentricity remaining within the signals. 
In order to be prepared to detect and study these signals, we will need eccentric \nr{} simulations and eccentric \gw{} models. 
In fact, some \lvk{} events have already shown potential evidence for eccentricity, increasing the urgency with which we need to explore this parameter space~\cite{Gayathri:2020coq, Romero-Shaw:2021ual}.
As part of this catalog, we have included 92 new eccentric simulations. 

Eighty of these simulations were part of a systematic suite designed to explore the eccentric space.
They have mass ratios of $1 \le q \le 4$ by steps of 1, and for each $q$, we have $a_1 = a_2 = \{0, 0.4\}$ aligned with the orbital angular momentum.
For each of the above cases, we performed simulations with eccentricity, $0.01 \le e \le 0.1$ by steps of $0.01.$
These simulations have catalog IDs of MAYA0913-MAYA0931, MAYA0938-MAYA0947, MAYA0949-MAYA0958, MAYA0960-MAYA0969, MAYA0971-MAYA0980,  MAYA0982-MAYA1001, and MAYA1041.
For each $q$, $a_1$, $a_2$ combination, the catalog also includes a quasicircular simulation.
All of these simulations begin at a separation of $11M$ and have 48 points across the radius of the finest grids which are sized to be just larger than the initial \bh{s}, with the exception of MAYA0992 which has 40 points across the radius of the finest grid.

Since eccentricity is a Newtonian concept, there is no unambiguous definition of eccentricity within \gr{}.
The \mayawaves{} library currently uses the orbital frequency definition described in ~\cite{Ramos-Buades:2018azo}.
This definition has proved to be robust and consistent with other definitions of eccentricity.
For setting up the eccentric simulations, we use the method described in ~\cite{Gayathri:2020coq}, applying a Newtonian derived correction to the quasicircular momentum.
Since this definition is based on Newtonian assumptions, the resulting eccentricities are not exactly the same as the target eccentricities.
As we are aiming to generally fill out the eccentric space and do not need precise values of $e=0.01, \,0.02, $... , we do not apply iterative corrections to target the exact eccentricity values.

We also include five additional simulations with $q=1$ and eccentricities up to $e=0.6$.
These have catalog IDs of MAYA0932-MAYA0936.
These have initial separations varying from $12M$ to $24M$ to avoid directly plunging with the higher eccentricities and have 60 points across the radius of the initial \bh{s}.
Additionally, we have seven simulations with both eccentricity and precession, with IDs MAYA1043-MAYA1049.
These are particularly useful since there are currently no eccentric, precessing waveform models.
These simulations have initial separations of $11M$ and 48 points across the radius of the initial \bh{s}.

Since eccentricity accelerates the inspiral of \bbh{s}, the number of cycles for these simulations varies greatly from 1.8 to 26.1 with an average of 14.7 cycles.

\subsection{Secondary Spin}\label{sec:high_mass_ratio}
We performed a suite of simulations spanning mass ratios of $q=1, 3, 5, 7, 15$ with varying spins, including nine simulations with $q=15$.
The primary \bh{} is nonspinning and the secondary \bh{} has an aligned spin of $-0.8 \le a_2 \le 0.8$ by steps of 0.2.
Current public \nr{} catalogs have minimal coverage of the parameter space with $q > 4$, especially with secondary spin.
These simulations help fill that gap.
The simulations with $q=1, 3, 5, 7$ begin with a separation of $D=11M$ and the $q=15$ simulations begin at $D=10M$.
The $q=1$ simulations have 36 points across the radius of the initial \bh{s} and the $q=3, 5, 7, 15$ simulations have 24 points across the radius of the initial \bh{s}.
Given the coarser grids of these simulations, parameter bias is likely to occur at lower \snr{s} ($\rho=14.7$ for the $q=5$ simulations and likely lower for the $q={7, 15}$ simulations).
While these waveforms are still useful for qualitative analysis, they are likely not high enough resolution for precision analyses.
The simulations in this suite range from having 11.4 \gw{} cycles up to 41.3 cycles with an average of 25.3 cycles. 
This suite includes simulations with IDs MAYA1002-MAYA1039.

\subsection{LVK Followup}\label{sec:lvk_followup}
Some of the detections made by the \lvk{} have required additional \nr{} followup.
This is particularly useful for events that push to the edges of the regions where models are sufficiently trained or in situations where different models yield inconsistent posteriors for the parameters of the binary.
We performed seven simulations (MAYA1050-MAYA1056) as followup for GW190521, an event that has been the focus of many studies due to its limited number of cycles~\cite{Gayathri:2020coq,Romero-Shaw:2021ual, LIGOScientific:2020iuh, Gamba:2021gap, CalderonBustillo:2020fyi}.
By using RIFT to perform parameter estimation on GW190521 with \nr{} simulations, these points in parameter space were identified as locations with both high likelihood and high uncertainty in the fit~\cite{Lange:2017wki}. 
Performing simulations at these points and then repeating the RIFT analysis with the new waveforms included is expected to improve the accuracy of and confidence in the estimated parameters for GW190521.
These simulations have an initial separation of $\sim 9.4M$, which was selected to have the provided initial parameters at a given initial frequency.
They are all set up to have 48 points across the radius of the initial \bh{s}.
Also for GW190521, we performed a simulation at the point of maximum likelihood for the purpose of visualizations for the announcement of the first intermediate-mass BH observation (MAYA0910).
We also performed one simulation (MAYA0912) as followup for GW170608.
The number of \gw{} cycles for the simulations in this suite range from 8.1 to 26.9 with an average of 10.9.

\subsection{Optimized Template Placement}\label{sec:template_placement}
Many of our simulations are performed for specific studies and thus have very specific initial parameters.
However, in order to prepare for future \lvk{} runs and next-generation detectors, we also want to fill out our catalog in an efficient and strategic way.
To accomplish this, we use the neural network presented in ~\cite{Ferguson:2022qkz} and the method it describes to identify optimal parameters.
This method makes use of a neural network trained to predict the match between any two quasi-circular binary systems given their initial parameters.
Using this network, we search for the minimal match between potential simulations and those already existing in the catalog.
Given the non-smooth nature of the space, we use basin hopping to identify global minima. 

This tool has enabled us to identify and perform simulations in a way that maximally fills in gaps in our coverage. 
We have just begun to use this tool and have performed 31 simulations suggested by it, with IDs MAYA1057-MAYA1087.
These simulations have mass ratios up to $q=5$ and precessing spins up to magnitudes of $0.8$.
All of these simulations have an initial separation of $11M$ and 48 points across the radius of the initial \bh{s}.
They contain between 9.9 and 24.8 \gw{} cycles with an average of 15.6 cycles.
Current catalogs have dense coverage in low mass-ratio spaces, but have minimal coverage above $q=4$.
In general, the precessing space is also much less densely covered than the aligned spin space.
These 31 simulations make significant improvements to the coverage of the moderate to high mass-ratio, precessing parameter space.
This tool will play a large role in the generation of waveforms for our future catalogs.

\section{Waveform Comparisons}\label{sec:model_comparisons}
Within the \nr{} community, there are several different methods and techniques for solving Einstein's equations directly. 
While each method has its benefits and drawbacks, all \nr{} waveforms are considered to the be the gold standard for studying the merger of comparable-mass \bbh{} systems.
To ensure accuracy and consistency in analyses that use \nr{} waveforms of each technique, it is important to perform cross-code comparisons.
Several of these have been done over the past couple decades, and the methods used by the \etk{} and the \sxs{} have proven to be reliable~\cite{2012CQGra..29l4001A, 2013CQGra..31b5012H, Healy:2017xwx}.
Here, we provide a comparison between this catalog and comparable waveforms available in the \sxs{} catalog. 

Similarly, there are many analytic and semi-analytic models for computing \gw{s}. 
These models have the benefit of being faster than \nr{} allowing for continuous sampling of the parameter space.
However, given that these models are trained or calibrated to \nr{} waveforms, they are generally not considered to be as accurate as \nr{} waveforms.
In this section we will also perform a comparison between this catalog and waveforms generated using such models.

For each of these comparisons, we compute the mismatch between two waveforms $h_1$ and $h_2$ as a noise weighted inner product optimized over time $t$ and phase $\phi$:
\begin{equation}
\mathcal{MM} = 1 - \max_{t,\, \phi} \mathcal O[h_1, h_2]  \equiv  1 - \max_{t,\, \phi} \frac{\langle h_1|h_2 \rangle}{\sqrt{\langle h_1|h_1\rangle \langle h_2|h_2\rangle}},  \label{eq:overlap}
\end{equation}
where
\begin{equation}
\langle h_1|h_2 \rangle = 2\int_{f_{0}}^\infty\frac{h_1^*h_2 +h_1\,h_2^*}{S_n} df\,,
\end{equation}
with $S_n$ being the noise power spectral density and $*$ denoting the complex conjugate~\cite{Owen:1995tm, Cutler:1994ys}.
To obtain detector independent results, we use a flat noise curve for this analysis.
Mismatches are very sensitive to systematic effects such as the amount of tapering and the starting frequency, and we 
find that these effects can change the mismatches by up to $\sim 10^{-3}$.
For consistency and to reduce Gibbs oscillations, we consider only waveforms with more than ten inspiral cycles and taper six of them. 
We compute all mismatches using a total mass of $M_{tot} = 100 M_\odot$ and a starting frequency of 30 Hz.

\begin{figure*}
\centering
\begin{subfigure}[t]{.48\textwidth}
    \centering
   \includegraphics[width=.95\linewidth]{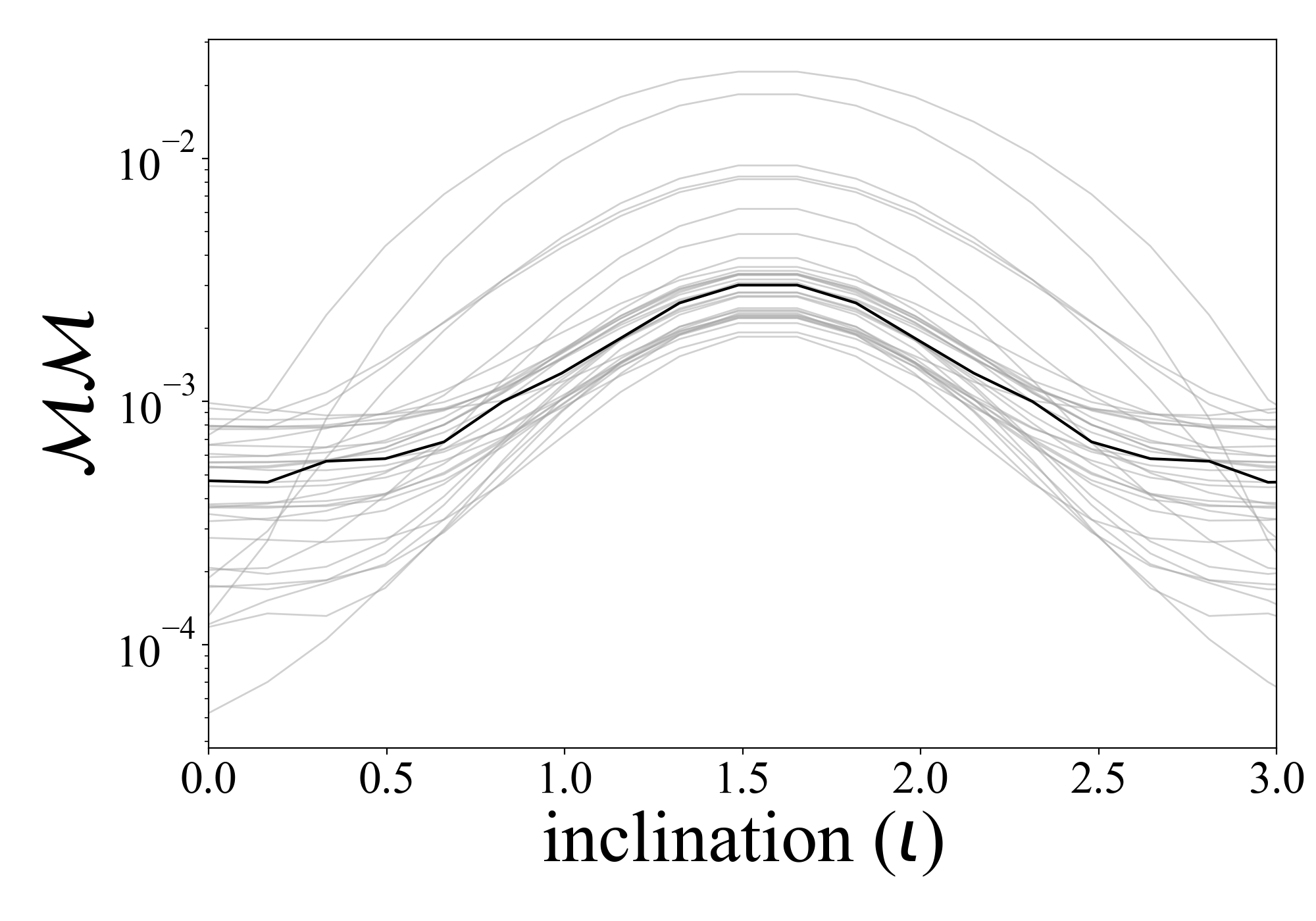}
    \caption{}
    \label{fig:sxs_mismatches_vs_inclination}
\end{subfigure}
\begin{subfigure}[t]{.48\textwidth}
    \centering
    \includegraphics[width=.95\linewidth]{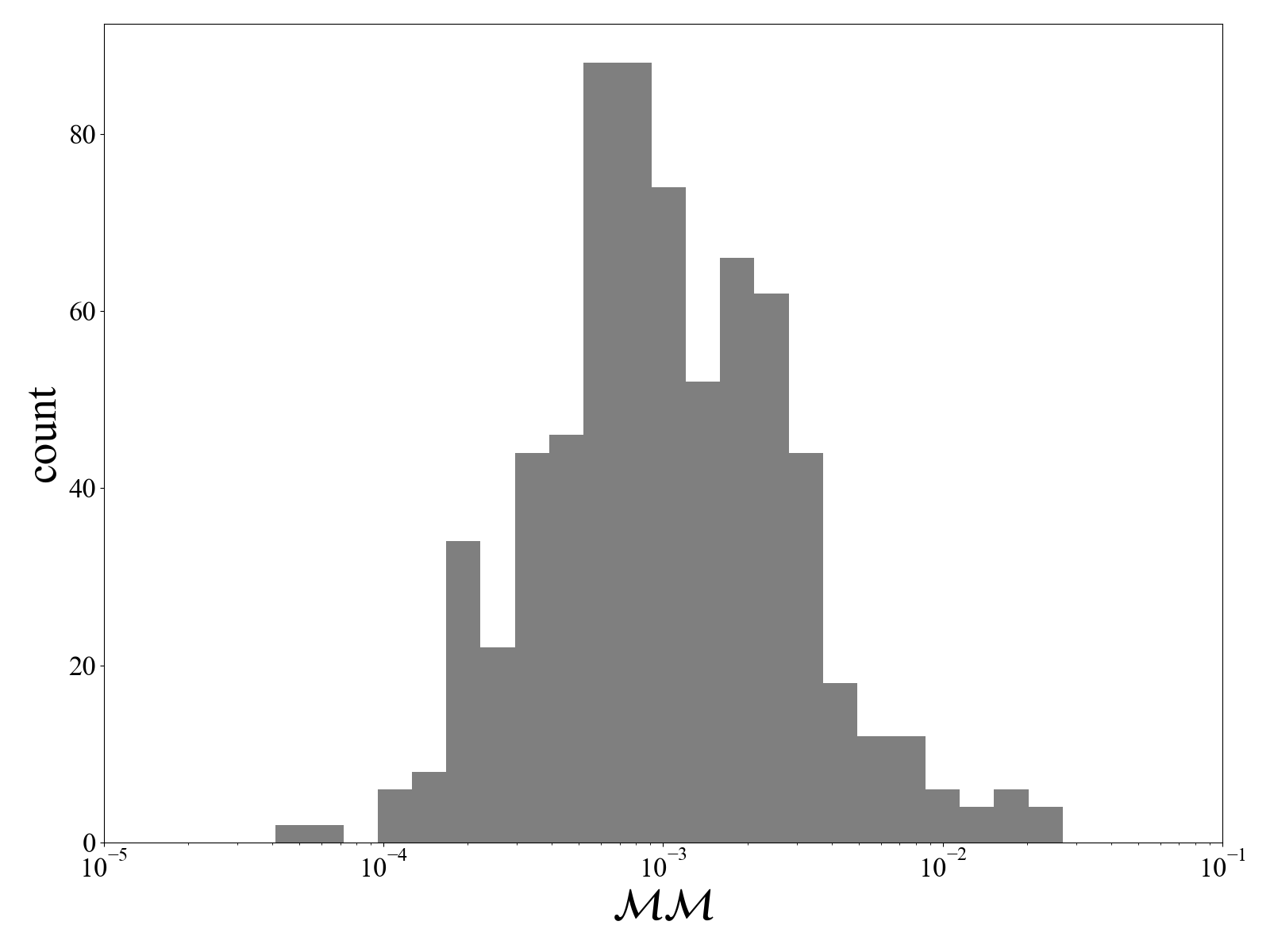}
 	\caption{}
 	\label{fig:sxs_mismatch_distribution}
\end{subfigure}
\caption{Mismatches between quasicircular, non-precessing MAYA waveforms and SXS waveforms using a flat noise curve over 20 different values of inclination. a) Mismatches as a function of inclination.  The black line is the median for all systems. The faint lines show all of the individual systems.  b) Distribution of mismatches.} 
\label{fig:sxs_mismatches}
\end{figure*}

For the \sxs{} comparison, we consider all equivalent non-precessing systems that exist in our catalog and in the \sxs{} catalog. 
Figure ~\ref{fig:sxs_mismatches_vs_inclination} shows the mismatch as a function of inclination, using all available modes and a flat noise curve. 
The grey lines show each individual system and the black line shows the median. 
Figure ~\ref{fig:sxs_mismatch_distribution} shows the distribution of mismatches (including all 20 inclinations) with a median mismatch of $9 \times 10^{-4}$.
The mismatches are consistently below $10^{-3}$ for face-on configurations with a median of $4.7 \times 10^{-4}$.
They increase with inclination, reaching a maximum for edge-on cases.
This is expected as higher modes become more prominent for higher inclination and are harder to resolve and more prone to systematics.
Some of the mismatch between \maya{} and \sxs{} can be attributed to the fact that the \maya{} waveforms considered for this analysis are at finite radius while the \sxs{} waveforms are extrapolated to infinite radius.

\begin{figure*}
\centering
\begin{subfigure}[t]{.48\textwidth}
    \centering
   \includegraphics[width=.95\linewidth]{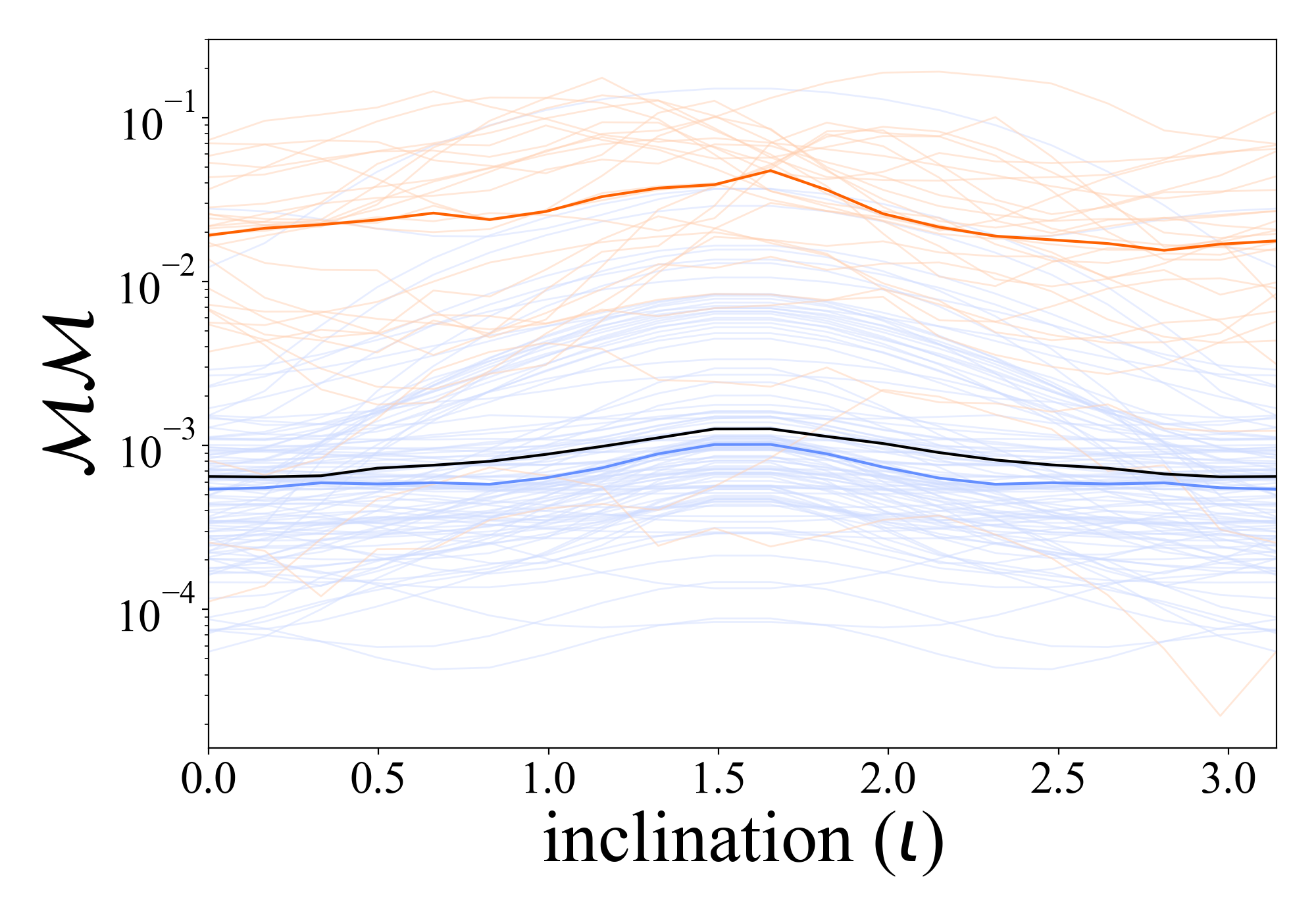}
    \caption{}
    \label{fig:mismatches_vs_inclination}
\end{subfigure}
\begin{subfigure}[t]{.48\textwidth}
    \centering
    \includegraphics[width=.95\linewidth]{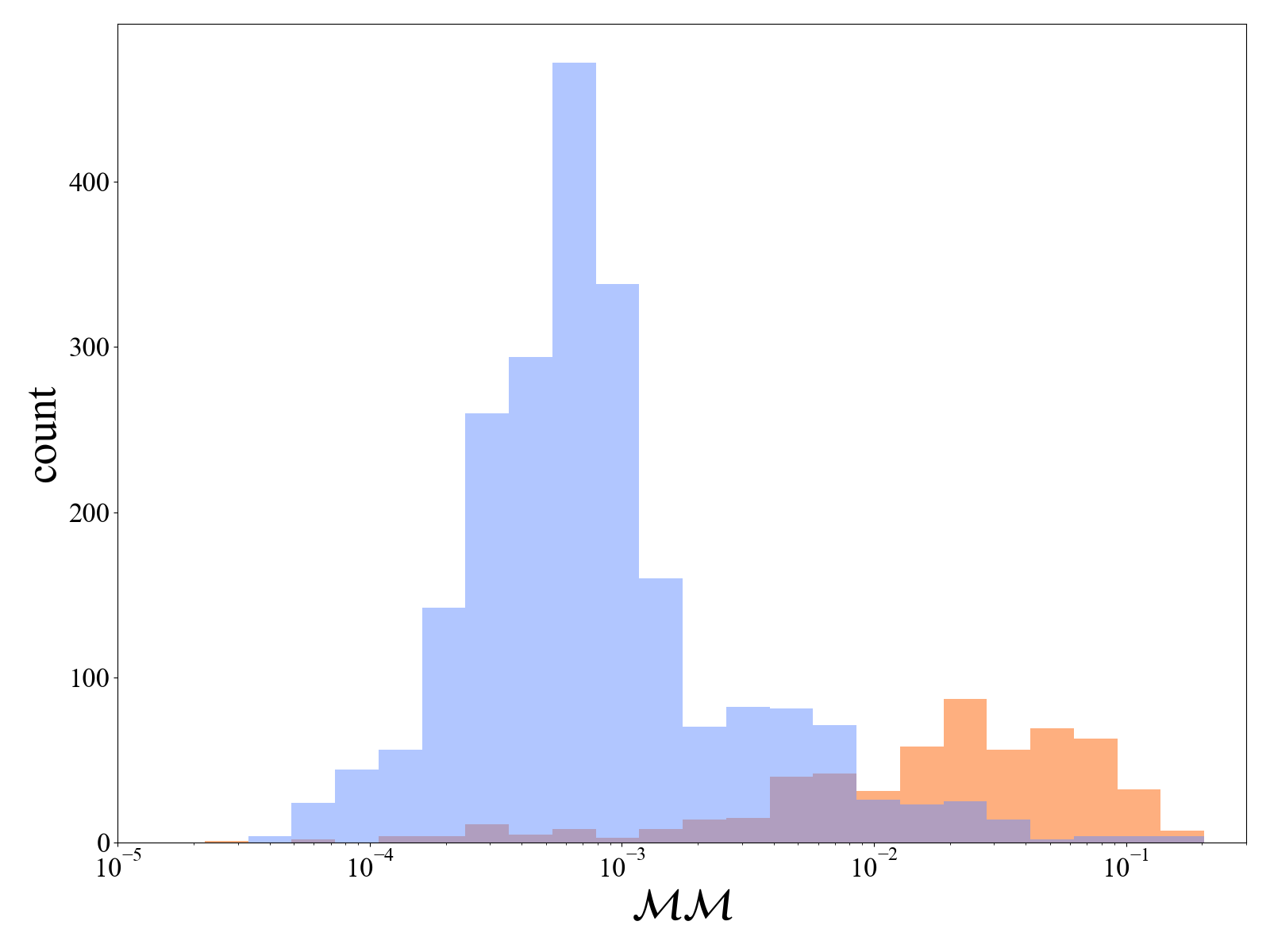}
 	\caption{}
 	\label{fig:mismatch_distribution}
\end{subfigure}
\caption{Mismatches between quasicircular MAYA waveforms and IMRPhenomXPHM waveforms using a flat noise curve over 20 different values of inclination. Orange denotes precessing systems and blue denotes non-precessing systems.  a) Mismatches as a function of inclination.  The bright orange and blue lines show the median of the precessing and non-precessing systems respectively.  The black line is the median for all included waveforms. The faint lines show all of the individual systems.  b) Distribution of mismatches. } 
\label{fig:model_mismatches}
\end{figure*}

We also compare our waveforms to the current state-of-the-art waveform models.
For the quasicircular case, we compare to IMRPhenomXPHM~\cite{Pratten:2020ceb} for both precessing and non-precessing systems, and for the eccentric case, we compare to TEOBResumS-DALI~\cite{Chiaramello:2020ehz, Nagar:2021gss, Nagar:2021xnh, Placidi:2021rkh}. 
Figure ~\ref{fig:mismatches_vs_inclination} shows the mismatch between the quasicircular \nr{} waveforms and the IMRPhenomXPHM waveforms as a function of inclination, including all available modes on a flat noise curve. 
These mismatches are minimized over in-plane spin angle as well as polarization angle.
Each of the precessing systems appears as a faint orange line with the vibrant orange line showing the median value for all precessing simulations.
All non-precessing simulations appear as faint blue lines with the vibrant blue line showing the median value for all non-precessing simulations. 
The black line shows the median mismatch of all quasicircular simulations.
Figure ~\ref{fig:mismatch_distribution} shows the distribution of the mismatches for precessing (orange) and aligned (blue) systems including 20 uniformly distributed inclinations.
For non-precessing systems, the median mismatch is $\sim 6.5 \times 10^{-4}$, and for the precessing systems, the median mismatch is $\sim 2.2 \times 10^{-2}$.

\begin{figure}
  \centering
  \includegraphics[width = 0.5 \textwidth]{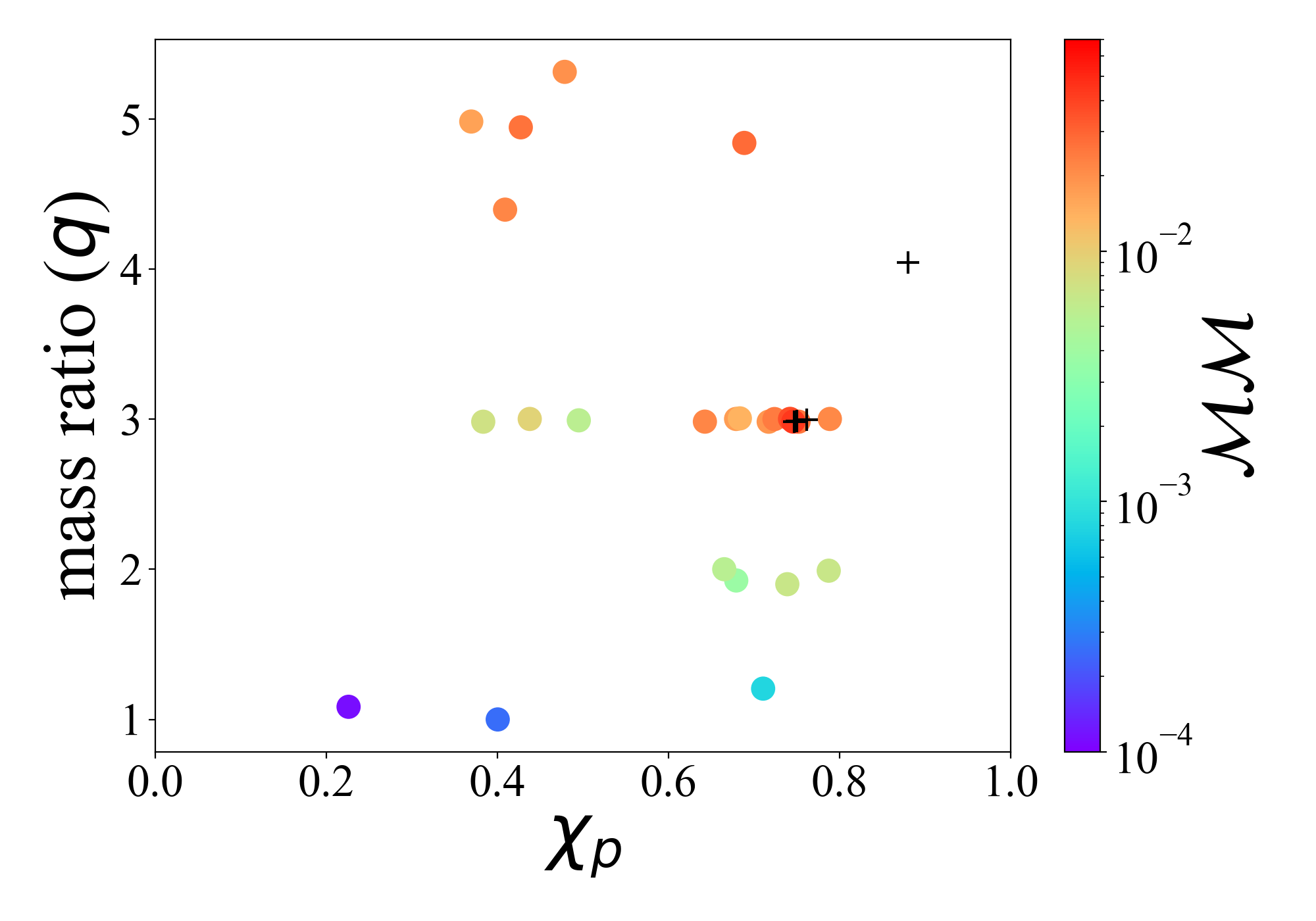}
  \caption{Mismatches between quasicircular MAYA waveforms and IMRPhenomXPHM waveforms in a face-on orientation as a function of the precessing spin parameter ($\chi_p$) and mass ratio ($q$).  The black crosses denote the four waveforms that have a mismatch greater than $0.05$.}
  \label{fig:mismatches_vs_chi_p}
\end{figure}

To better understand the higher precessing mismatches, in Figure ~\ref{fig:mismatches_vs_chi_p} we plot the mismatch as a function of the mass ratio and the precessing spin parameter ($\chi_p$)~\cite{Schmidt:2012rh} where
\begin{equation}
\chi_{p} = max\left(a_1 sin(\theta_1), \frac{4 m_2 ^ 2 + 3 m_1 m_2}{4 m_1 ^2 + 3 m_1 m_2} a_2 sin(\theta_2) \right) \, ,
\end{equation}
with $a_{1,2}$ being the dimensionless spin parameters of the primary and secondary \bh{s} respectively and $\theta_{1,2}$ being the angle of those spins with respect to the direction of the orbital angular momentum.
It can be seen that the mismatch increases with mass ratio and $\chi_p$, consistent with the fact that such regions of parameter space are more challenging to model.
The four simulations marked with black crosses have mismatches greater than $0.05$.
Each of these simulations has $\chi_p \geq 0.75$ and $q \geq 3$.
This is consistent with the mismatches presented in the IMRPhenomXPHM paper for such systems.

\begin{figure*}
\centering
\begin{subfigure}[t]{.48\textwidth}
    \centering
   \includegraphics[width=.95\linewidth]{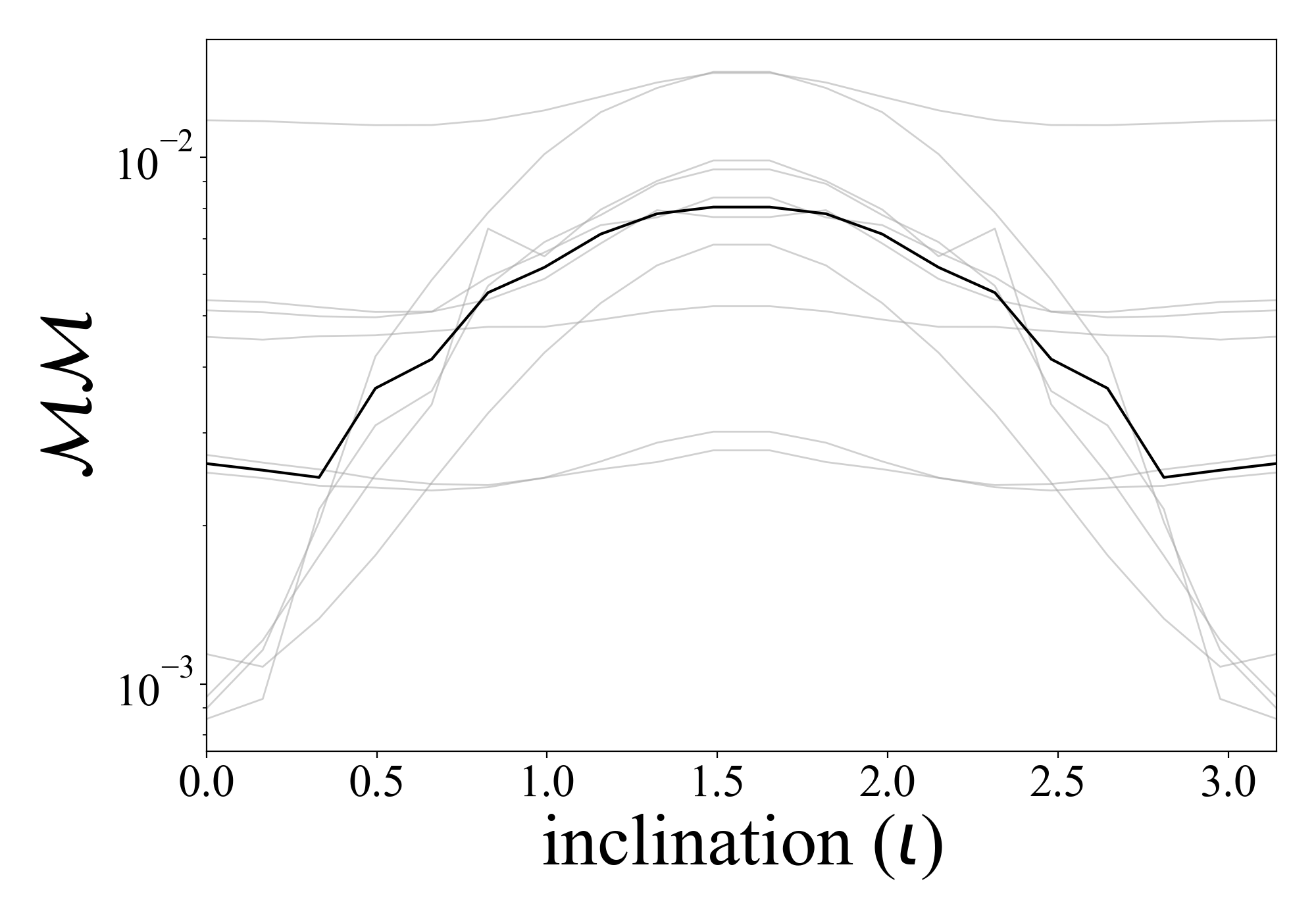}
    \caption{}
    \label{fig:eccentric_mismatches_vs_inclination}
\end{subfigure}
\begin{subfigure}[t]{.48\textwidth}
    \centering
    \includegraphics[width=.95\linewidth]{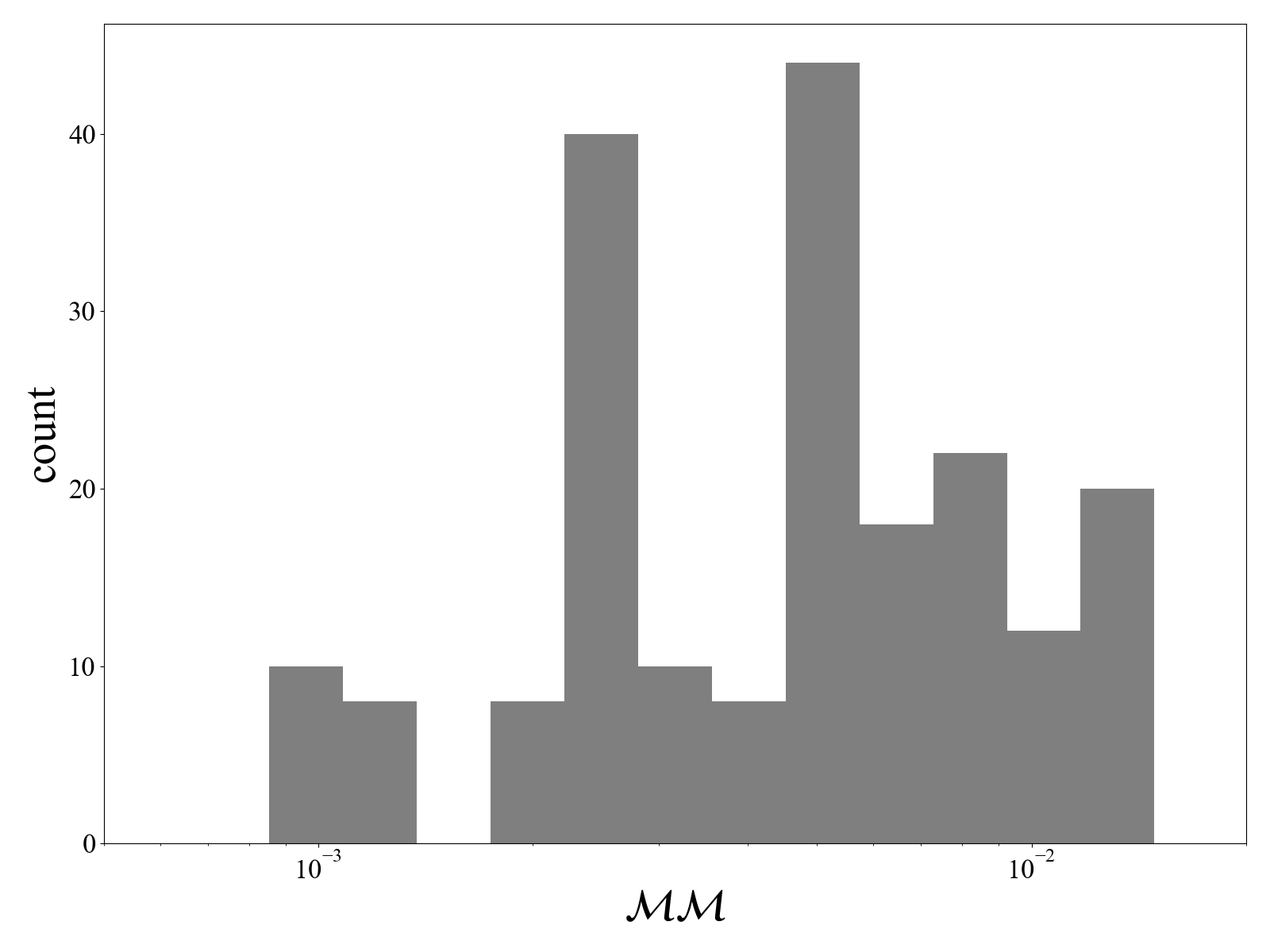}
 	\caption{}
 	\label{fig:eccentric_mismatch_distribution}
\end{subfigure}
\caption{Mismatches between eccentric, non-precessing MAYA waveforms and TEOBResumS-DALI waveforms using a flat noise curve over 20 different values of inclination. a) Mismatches as a function of inclination.  The black line is the median for all systems. The faint lines show all of the individual systems.  b) Distribution of mismatches.} 
\label{fig:eccentric_mismatches}
\end{figure*}

For the eccentric case, we use TEOBResumS-DALI as the waveform model and consider only aligned spin systems.
We optimize over the initial eccentricity and frequency of the model waveforms to obtain the best mismatch, since the model's eccentricity parameter is not physically meaningful.
Figure ~\ref{fig:eccentric_mismatches} shows the mismatches between TEOBResumS-DALI waveforms and ten of our waveforms which fall within the validity range of the model ($e \leq 0.2$,  $q \leq 3$,  and $a_1, a_2 \leq 0.7$). 
The median mismatch over all systems and inclinations is $\sim 5 \times 10^{-3}$.

The mismatches between the MAYA Catalog and other NR catalogs and waveform models average on order $10^{-3}$, showing that the MAYA Catalog is consistent with other state-of-the-art waveforms.
It also shows that waveform models are sufficiently capturing the gravitational radiation accurately, though for higher mass ratios and strong precession, the models become less accurate.

\section{Final Mass, Spin, and Kick}\label{sec:final_state}

In this section, we compare the properties of the remnant \bh{} computed from \nr{} simulations to those predicted using NRSur7dq4Remnant~\cite{Varma:2019csw} for quasi-circular binaries.
We use the SurfinBH python library ~\cite{Varma:2018aht, vijay_varma_2018_1435832} to interact with NRSur7dq4Remnant and compute the predicted remnant properties.
We provide the masses and spins of the initial \bh{s} at a time $100 M$ before merger, rotated into the same frame used by NRSur7dq4Remnant.
We only consider systems which fall within the range of validity for NRSur7dq4Remnant ($q \leq 4$ and dimensionless spin parameters $a_1, a_2 \leq 0.8$). 
The masses and spins of the initial and remnant \bh{s} are computed from the \bh{} apparent horizons, and the magnitude of the kick velocity is computed from the linear momentum radiated by the \gw{s}.

\begin{figure*}
\centering
\begin{subfigure}{.3\textwidth}
    \centering
   \includegraphics[width=.95\linewidth]{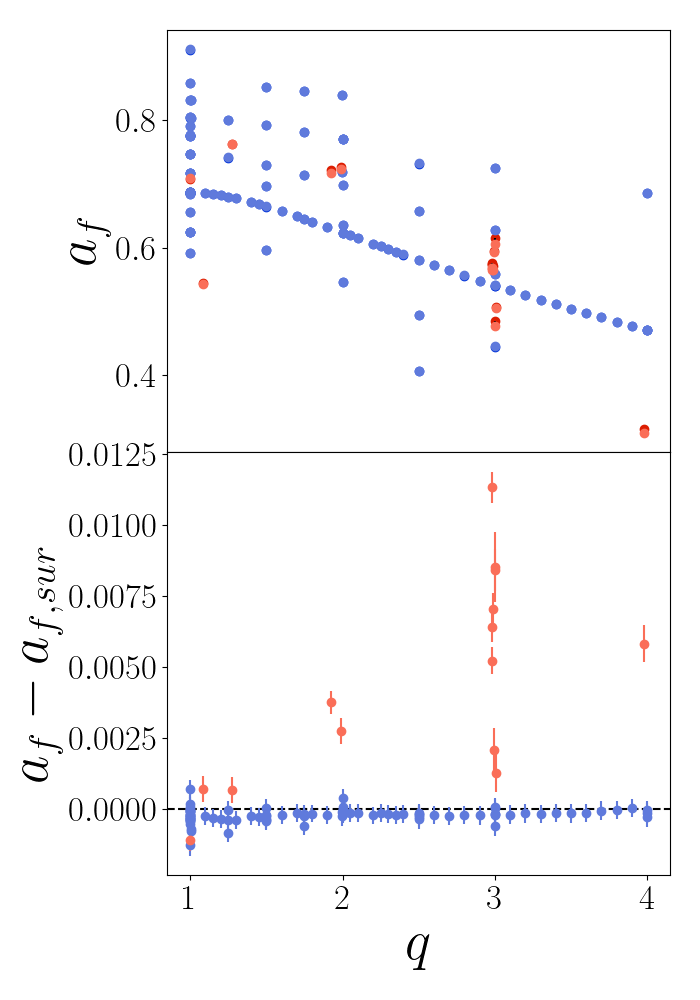}
    \caption{ }
    \label{fig:remnant_af}
\end{subfigure}
\begin{subfigure}{.3\textwidth}
    \centering
    \includegraphics[width=.95\linewidth]{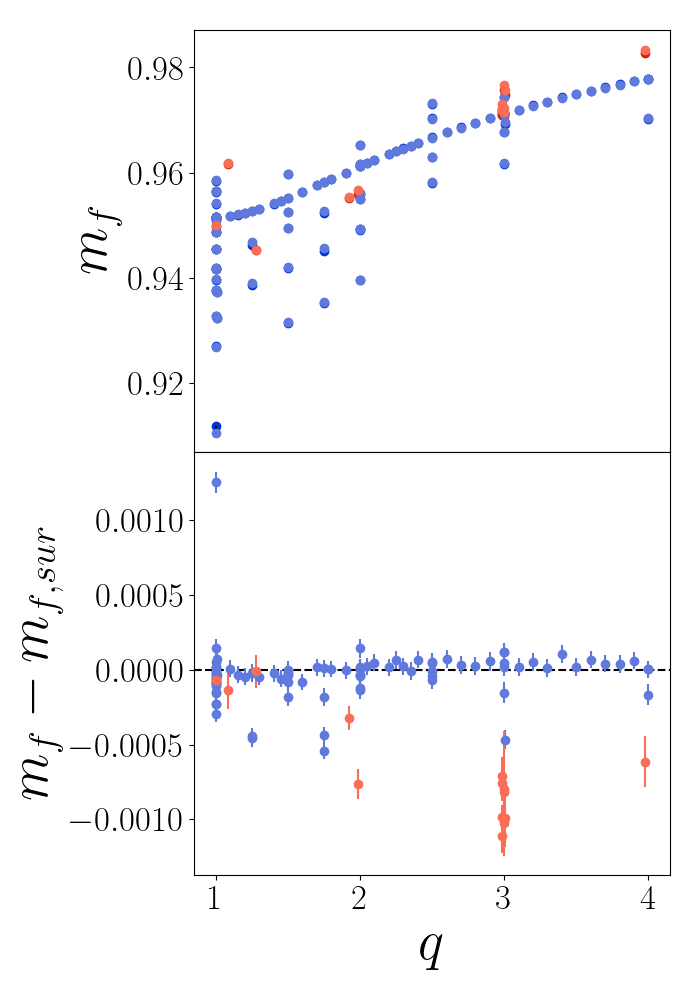}
 	\caption{ }
 	\label{fig:remnant_mf}
\end{subfigure}
\begin{subfigure}{.3\textwidth}
    \centering
    \includegraphics[width=.95\linewidth]{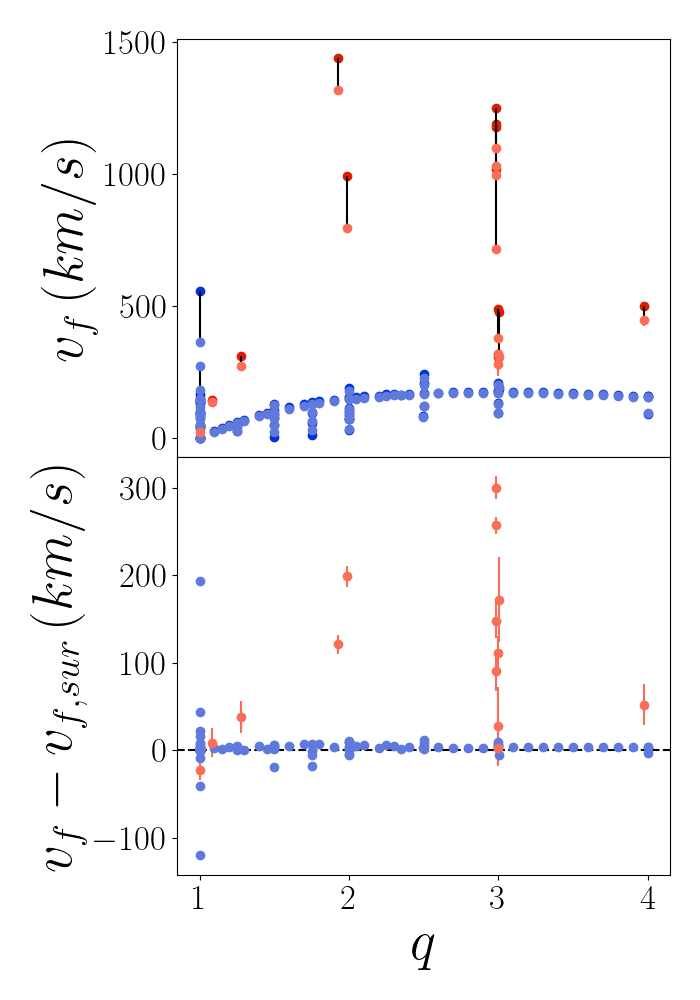}
 	\caption{ }
 	\label{fig:remnant_vf}
\end{subfigure}
\caption{Comparison of the (a) spin magnitude, (b) mass, and (c) recoil speed of the remnant black hole computed from the NR simulation and predicted by NRSur7dq4Remnant. The top panels show the remnant property as a function of mass ratio.  Non-precessing systems appear in blue and precessing systems appear in orange. The darker points show the \nr{} values and the lighter points show the model predictions. The points referring to the same systems are connected by black lines.  The bottom panel shows the residuals resulting from subtracting the model predicted value from the \nr{} value. The vertical lines are the uncertainty in the model predictions.}
\label{fig:final_state}
\end{figure*}

Figure ~\ref{fig:final_state} shows the remnant properties obtained from \nr{} simulations compared to the remnant properties obtained from NRSur7dq4Remnant.
For each of the remnant properties, the non-precessing systems have much lower errors than the precessing systems.
However, for remnant spin,  all residuals fall within $~0.0125$, or $2.55\%$, and for remnant mass, all residuals fall within $~0.0012$, or $0.14\%$.
The recoil velocity is more challenging since it is very sensitive to spin angles. 
The system with the largest residual has an error of $29\%$.

\section{Center of Mass Drift Corrections}\label{sec:com_correction}

During the \nr{} simulation of the  inspiral of \bbh{} systems, the center of mass experiences a wobble and a drift. 
These are caused by a combination of physical effects due to the asymmetrical emission of \gw{s} and numerical error.
Since the \gw{s} are decomposed with the spherical harmonics centered on the initial \com{}, this can cause mode mixing.
In recent years there has been much discussion of the impact that these \com{} drifts can cause to the waveform modes.
We include here a brief analysis of the impact of correcting for such drifts.

\begin{figure*}
\centering
\begin{subfigure}[t]{.48\textwidth}
    \centering
   \includegraphics[width=.95\linewidth]{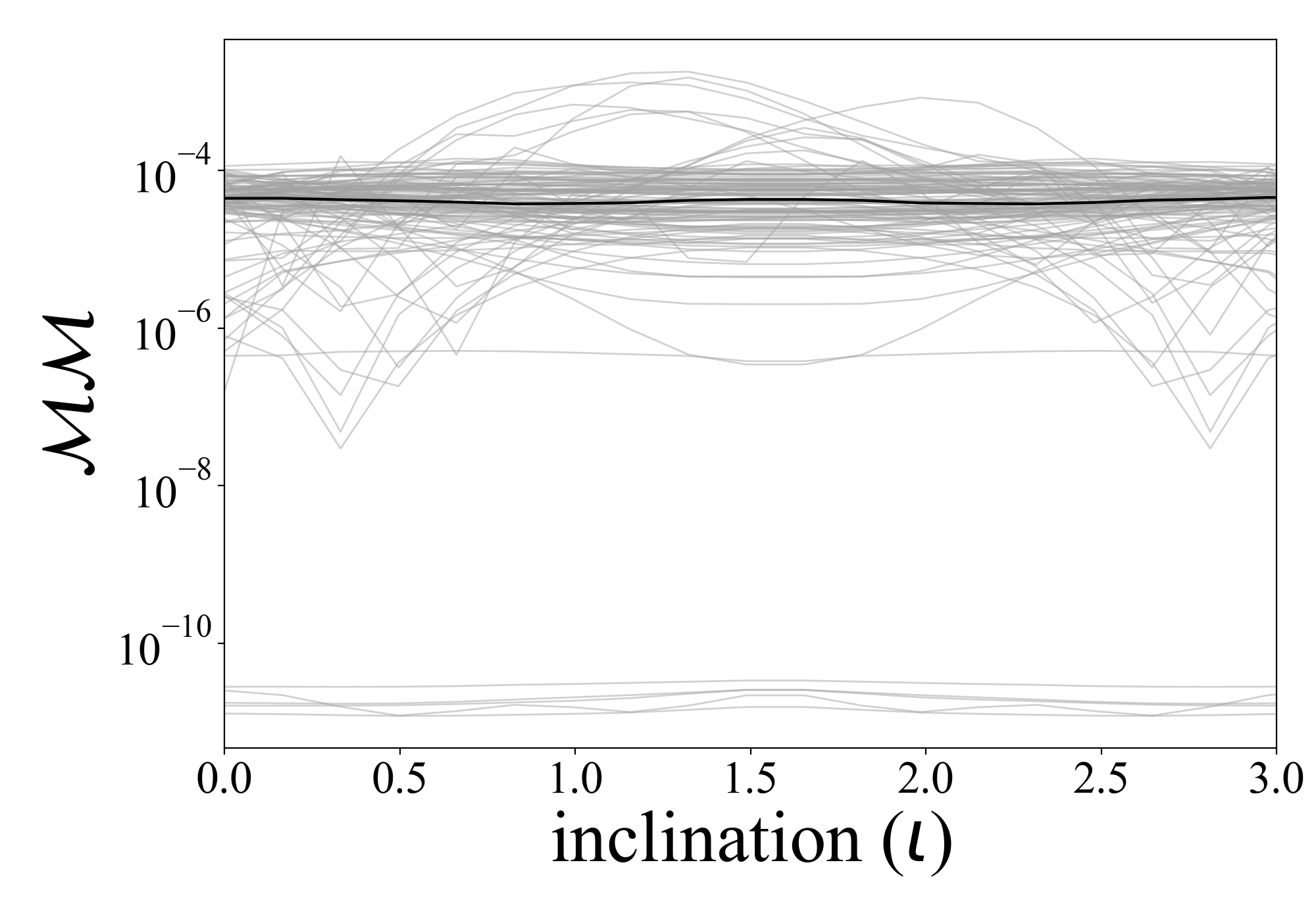}
    \caption{}
    \label{fig:com_mismatches_vs_inclination}
\end{subfigure}
\begin{subfigure}[t]{.48\textwidth}
    \centering
    \includegraphics[width=.95\linewidth]{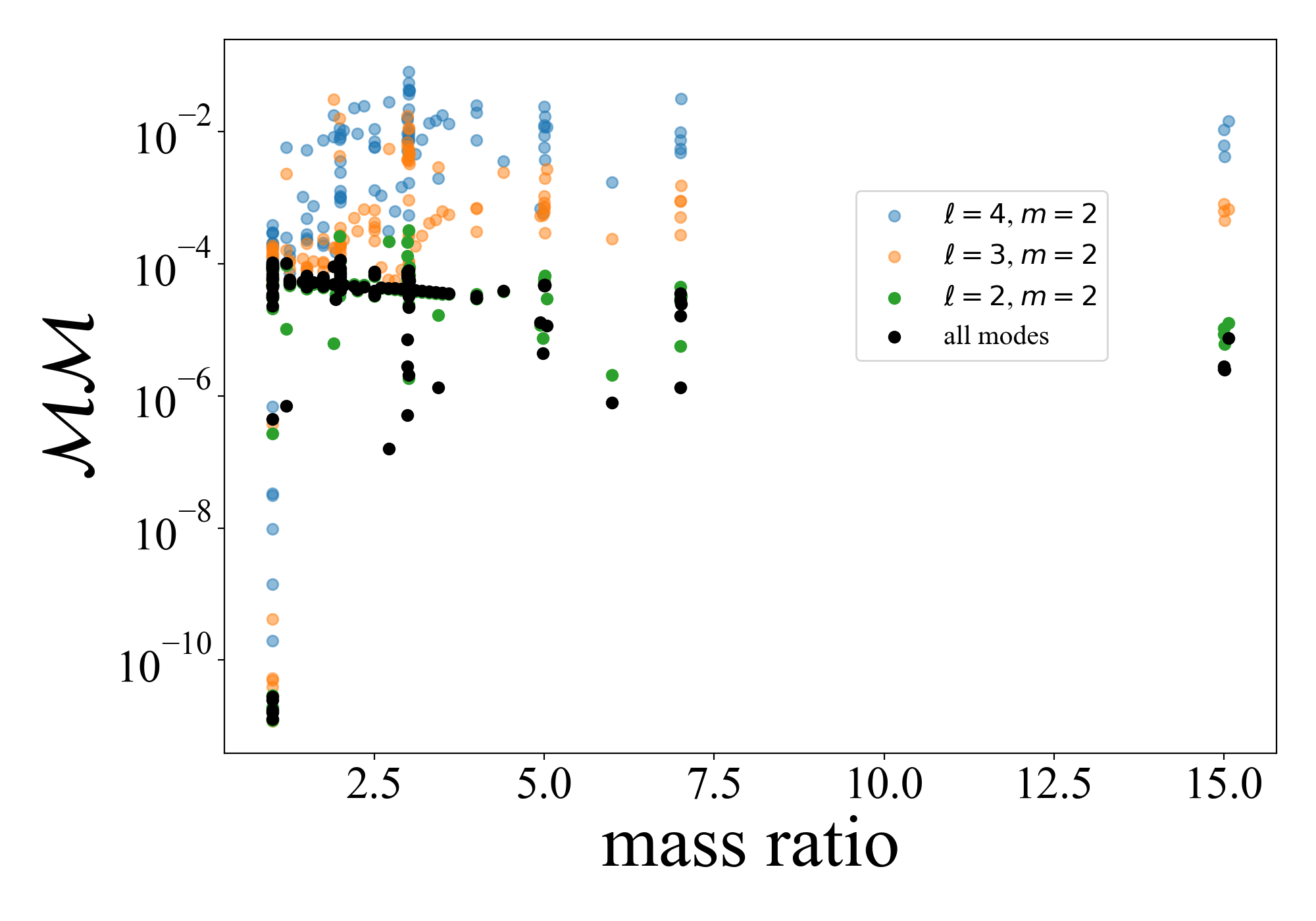}
 	\caption{}
 	\label{fig:com_mismatches_vs_mass_ratio}
\end{subfigure}
\caption{Mismatches between \com{} drift corrected waveforms and raw waveforms. a) The mismatch including all modes vs inclination. Each system is represented by a light gray line with the darker line showing the median value.  b) The mismatch for various modes and for all modes (at face-on orientation) as a function of mass ratio. } 
\label{fig:com_mismatches}
\end{figure*}

Using the Scri python package~\cite{mike_boyle_2020_4041972, Boyle2013, BoyleEtAl:2014,Boyle2015a}, we apply corrections for a translation and boost in the \com{}, computed using equations 6 and 7 of ~\cite{Woodford:2019tlo}.
We then perform a mismatch analysis comparing the \com{} corrected waveforms and the uncorrected waveforms for the overall strain and for individual modes.
Figure ~\ref{fig:com_mismatches} shows the mismatches between raw and \com{} corrected strains.
In Figure ~\ref{fig:com_mismatches_vs_inclination}, we see the mismatch with all modes combined as a function of inclination. 
The median mismatch is fairly constant for all inclinations, with a value of $~\sim 4 \times 10^{-5}$.
In Figure ~\ref{fig:com_mismatches_vs_mass_ratio}, we see the mismatch values for several modes as a function of mass ratio.
While there does not appear to be a strong correlation between the impact of \com{} corrections and mass ratio, the figure does reveal that certain modes are more impacted than others.
In particular,  the $\ell = 4$, $m = 2$ mode has mismatches up to $\sim 10^{-2}$.

Using the \mayawaves{} library, our waveforms can be read in their raw form or with \com{} corrections. 
The waveforms stored in the \pycbc{} compatible format~\cite{lalsuite, alex_nitz_2020_3630601, Schmidt:2017btt} have been corrected for \com{} drift.
Based on these results, we would recommend using the \com{} corrections for mode-by-mode analyses.
For analyses using all the modes recombined, the corrections aren't as significant but could still be beneficial.

\section{Conclusion}\label{sec:conclusion}
This paper introduces the second MAYA catalog of \nr{} waveforms. 
It contains 181 waveforms that fill in and expand our coverage of the \bbh{} parameter space.
This includes 80 from an eccentric suite ranging up to eccentricities of 0.1, as well as seven systems with both precession and eccentricity.
It also includes 38 waveforms studying secondary spin, 9 simulations as \nr{} followup for \lvk{} events, and 31 simulations placed to optimize parameter space coverage.

As the most accurate way of studying the merger of comparably massed objects, \nr{} waveforms are crucial for understanding the observations of current and next-generation \gw{} detectors.
We must, therefore, ensure we have sufficient \nr{} coverage of the anticipated parameter space.
This catalog makes significant strides towards this goal by pushing and filling in considerable gaps in existing \nr{} coverage of the \bbh{} parameter space. 
With mass ratios up to $q=15$, highly precessing simulations, and an extensive eccentric suite, this catalog greatly improves the coverage of public \nr{} catalogs.

By performing a thorough error analysis and comparison to other \nr{} codes, we are confident in the accuracy and reliability of the waveforms presented in this catalog.
Through a comparison to state-of-the-art \gw{} models, we show consistency but also highlight the continuing need for \nr{} waveforms particularly in the less well understood precessing space. 

The waveforms are provided at \url{https://cgpstorage.ph.utexas.edu/waveforms} and can be read using the \mayawaves{} python package. 
We have also updated the previous Georgia Tech catalog to include more data by using the new format, leading to a total of 635 public waveforms.
All waveforms that fit the requirements are also provided in the format described in ~\cite{Schmidt:2017btt} for use with \pycbc{}. \\

\paragraph*{\textbf{Acknowledgements}}
The catalog and analysis presented in this paper are possible due to grants NASA 80NSSC21K0900, NSF 2207780, NSF 2114582, and NSF OAC-2004879. 
The computing resources necessary to perform the simulations in this catalog were provided by XSEDE PHY120016, TACC PHY20039, TACC NR-Catalog and TACC Template-Placement.
This work was done by members of the Weinberg Institute and has an identifier of UTWI-32-2023.

\appendix

\section{Convergence Testing}\label{sec:convergence}

For finite differencing codes, we can relate $\Psi_4$ for a finite resolution simulation to the $\Psi_4$ for an infinite resolution simulation according to the following:
\begin{equation}
\Psi_{4,i} = \Psi_4 + c \Delta_i^\alpha \, ,
\end{equation}
where $i$ denotes a given resolution, $c$ is a function of time independent of resolution, $\Delta_i$ is the grid spacing on the finest grid, and $\alpha$ is the convergence rate of the code.
Using this expression for three resolutions, we obtain the following:
\begin{equation}
\frac{\Psi_{4,high} - \Psi_{4, med} }{\Psi_{4,med} - \Psi_{4, low}} = \frac{\Delta_{high}^{\alpha} - \Delta_{med}^{\alpha}}{\Delta_{med}^{\alpha} - \Delta_{low}^{\alpha}} \, ,
\end{equation}

\begin{figure*}
\centering
\begin{subfigure}{.45\textwidth}
    \centering
    \includegraphics[width=\linewidth]{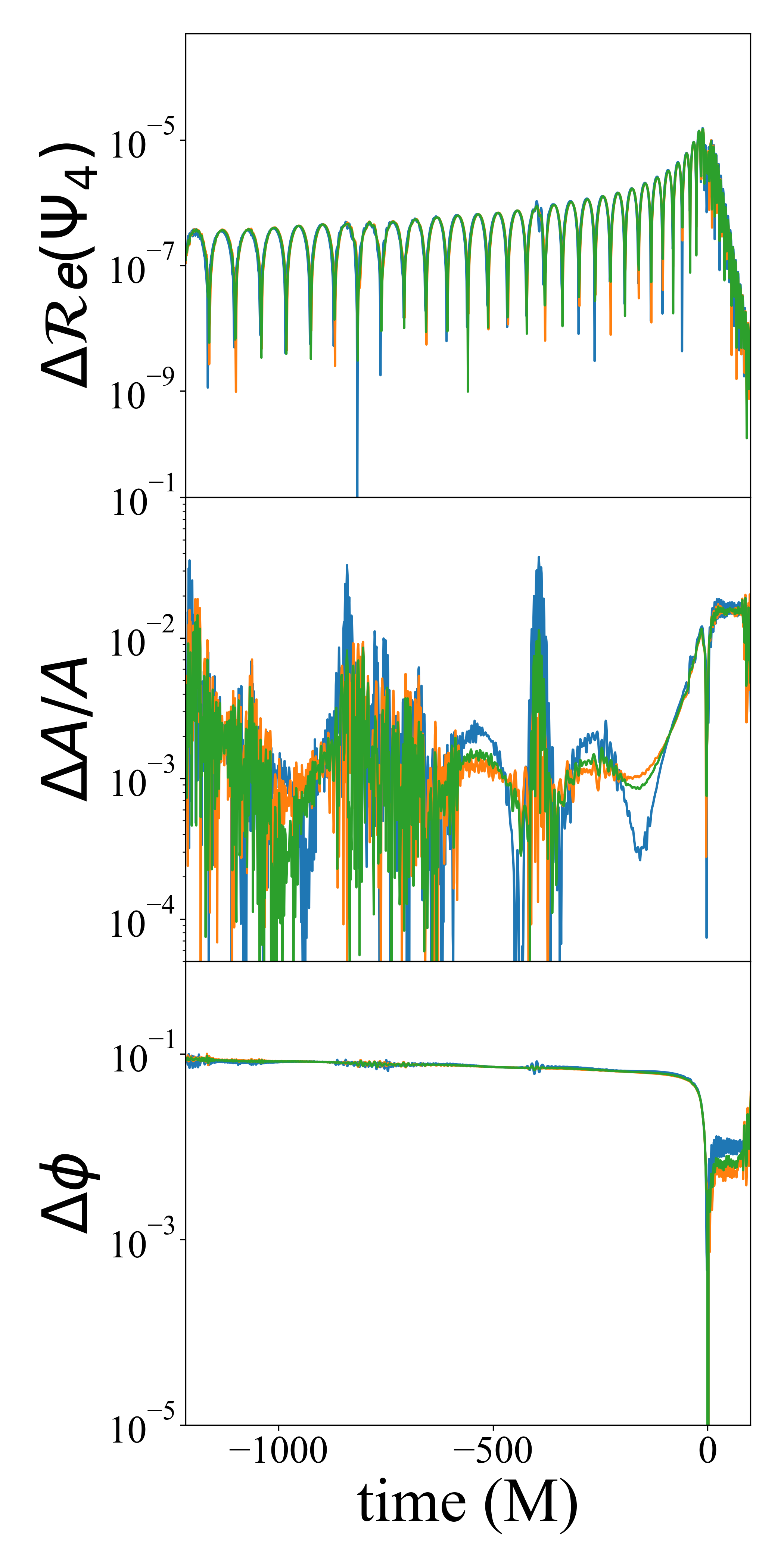}
\caption{}
\label{fig:convergence_q1}
\end{subfigure}
\begin{subfigure}{.45\textwidth}
    \centering
    \includegraphics[width=\linewidth]{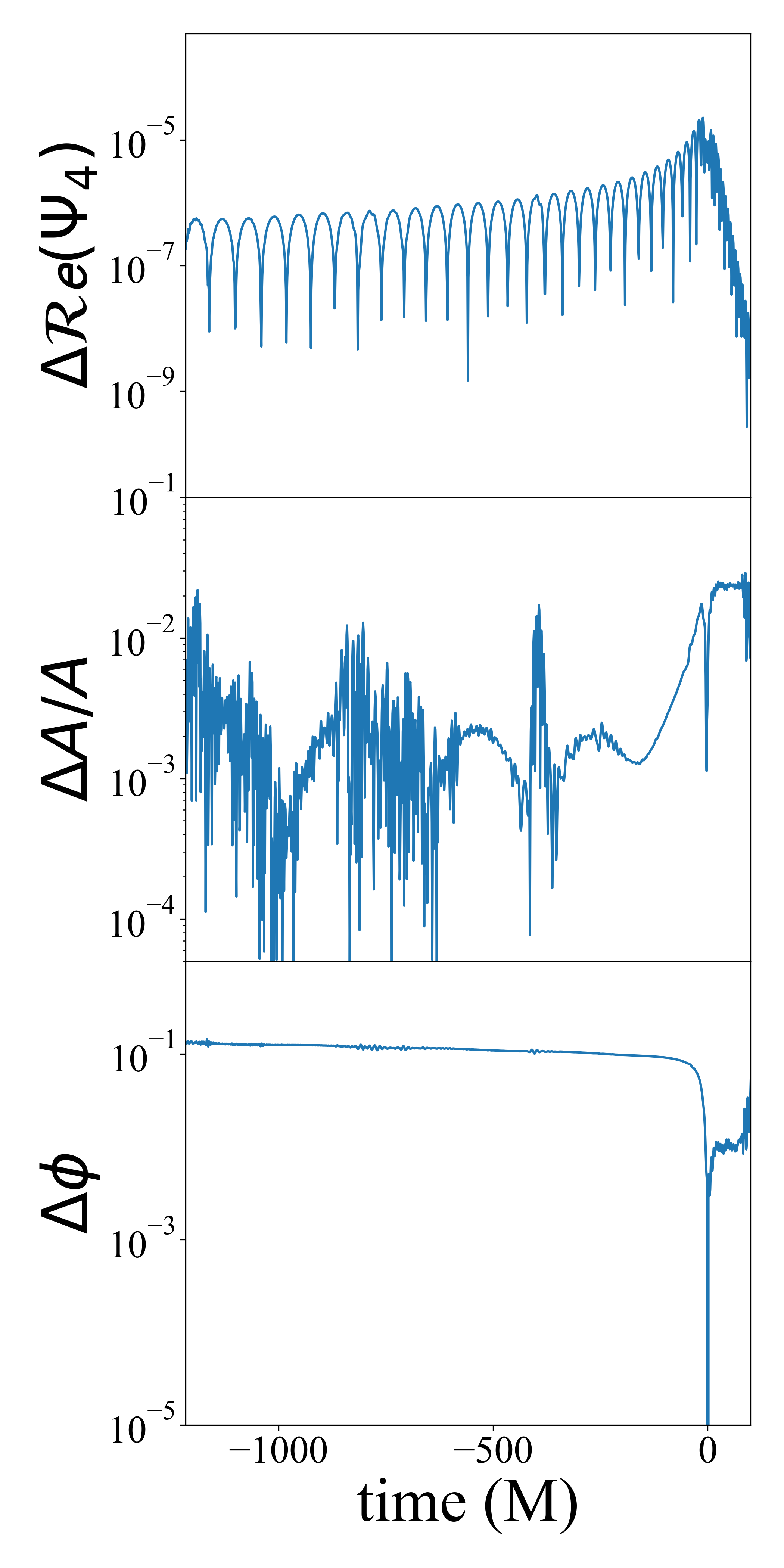}
\caption{}
\label{fig:est_error_q1}
\end{subfigure}
\caption{Convergence and error analysis for an equal mass, nonspinning, quasi-circular simulation of separation 11M.  a) The residuals for the real part, amplitude, and phase of $\Psi_4$ using the high and medium resolutions (blue), the medium and low resolutions (orange), and the high and low resolutions (green). The orange and green lines are scaled according to the convergence factor using Eq. ~\ref{eq:convergence_function}. b) Estimated error in the standard resolution waveform.}
\label{fig:q1}
\end{figure*}

To verify the convergence of our code, we perform an equal mass, nonspinning simulation with separation 11M (consistent with the separation of most of our simulations) at three different resolutions: $M/123.08$, $M/184.62$, and $M/276.92$. 
These correspond to 32, 48, and 72 points across the radius of the \bh{} respectively.
The middle resolution is consistent with those presented in this catalog, as we typically set up our grids such that there are 48 points across the radius of each \bh{}.

We find the convergence rate by computing the residuals between the high and medium resolutions and the medium and low resolutions and finding the value of $\alpha$ such that they scale according to:
\begin{equation}
\Psi_{4,high} - \Psi_{4, med}  = a(\Psi_{4,med} - \Psi_{4, low}) \, ,
\end{equation}
where 
\begin{equation}
a=\frac{\Delta_{high}^{\alpha} - \Delta_{med}^{\alpha}}{\Delta_{med}^{\alpha} - \Delta_{low}^{\alpha}} \, .
\label{eq:convergence_function}
\end{equation}
The convergence rate proves to be $\alpha = 2.75$ for this simulation.
We use 6-th order spacial finite differencing and 4th order Runge-Kutta for time evolution. 
Given the complex mesh refinement and the interpolation necessary to extract $\Psi_4$ on spheres, this convergence rate is as expected. 
We verify this convergence rate, and confirm that the simulations are in the convergent regime, by showing that the residual between the high and low resolutions also scales according to the computed value of $\alpha$.
Figure ~\ref{fig:convergence_q1} shows the residual between the high and medium resolutions as well as the residuals between the medium and low resolutions and the high and low resolutions scaled by the associated factor.
The waveforms are aligned such that both time and phase at peak amplitude are 0.

We use the results from our convergence test to estimate the errors in our standard resolution waveform (the medium resolution waveform in this case).  
We do so by scaling the residual between two resolutions according to the factor described above to estimate the error between our standard resolution and infinite resolution:
\begin{equation}
\Psi_{4,inf} - \Psi_{4, med} = \frac{ - \Delta_{med}^{\alpha}}{\Delta_{high}^{\alpha} - \Delta_{med}^{\alpha}} * (\Psi_{4,high} - \Psi_{4, med})\, .
\end{equation}
Figure ~\ref{fig:est_error_q1} shows the estimated error in our standard resolution waveform for this system.
The amplitude error reaches $\sim 2.3\%$  at merger, and the phase error reaches $\sim 0.12$.

Then, using the convergence rate and the mismatches between the waveforms ($\sim 10^{-4}$ between the lowest and highest resolution waveforms), we can compute the critical \snr{} at which the finite resolution effects will become significant.
Based on the method presented in ~\cite{Ferguson:2020xnm}, our standard resolution waveform will be indistinguishable from an ``infinite'' resolution face-on signal up to an \snr{} of $\rho = 192$.
Based on the modified expression presented in ~\cite{Jan:2023raq}, this will be unlikely to cause significant parameter estimation bias up to $\rho = 543$ for an eight dimensional analysis (mass and spin vector for each \bh{}).

\begin{figure*}
\centering
\begin{subfigure}{.45\textwidth}
    \centering
    \includegraphics[width=\linewidth]{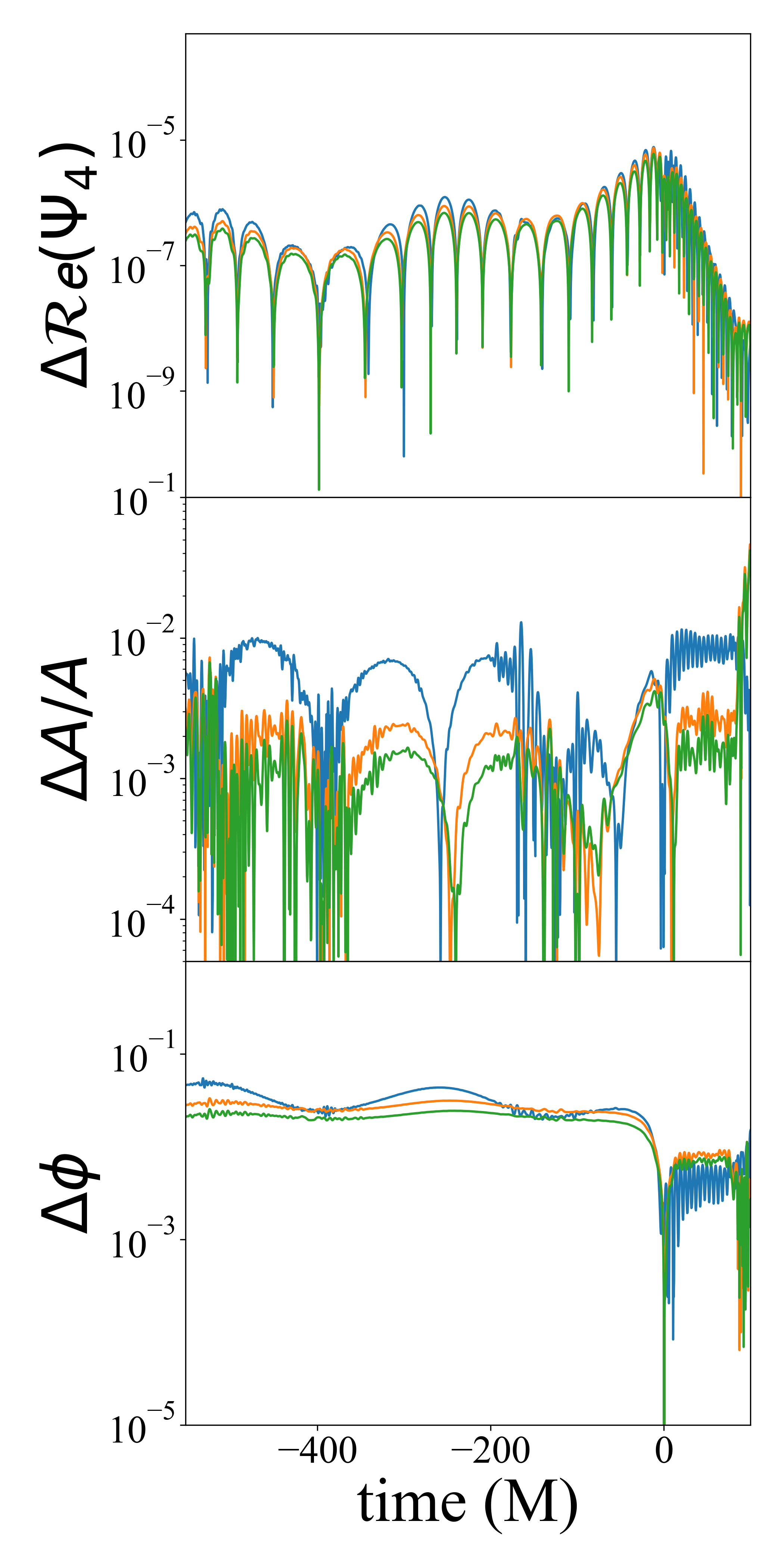}
\caption{}
\label{fig:convergence_eccentric}
\end{subfigure}
\begin{subfigure}{.45\textwidth}
    \centering
    \includegraphics[width=\linewidth]{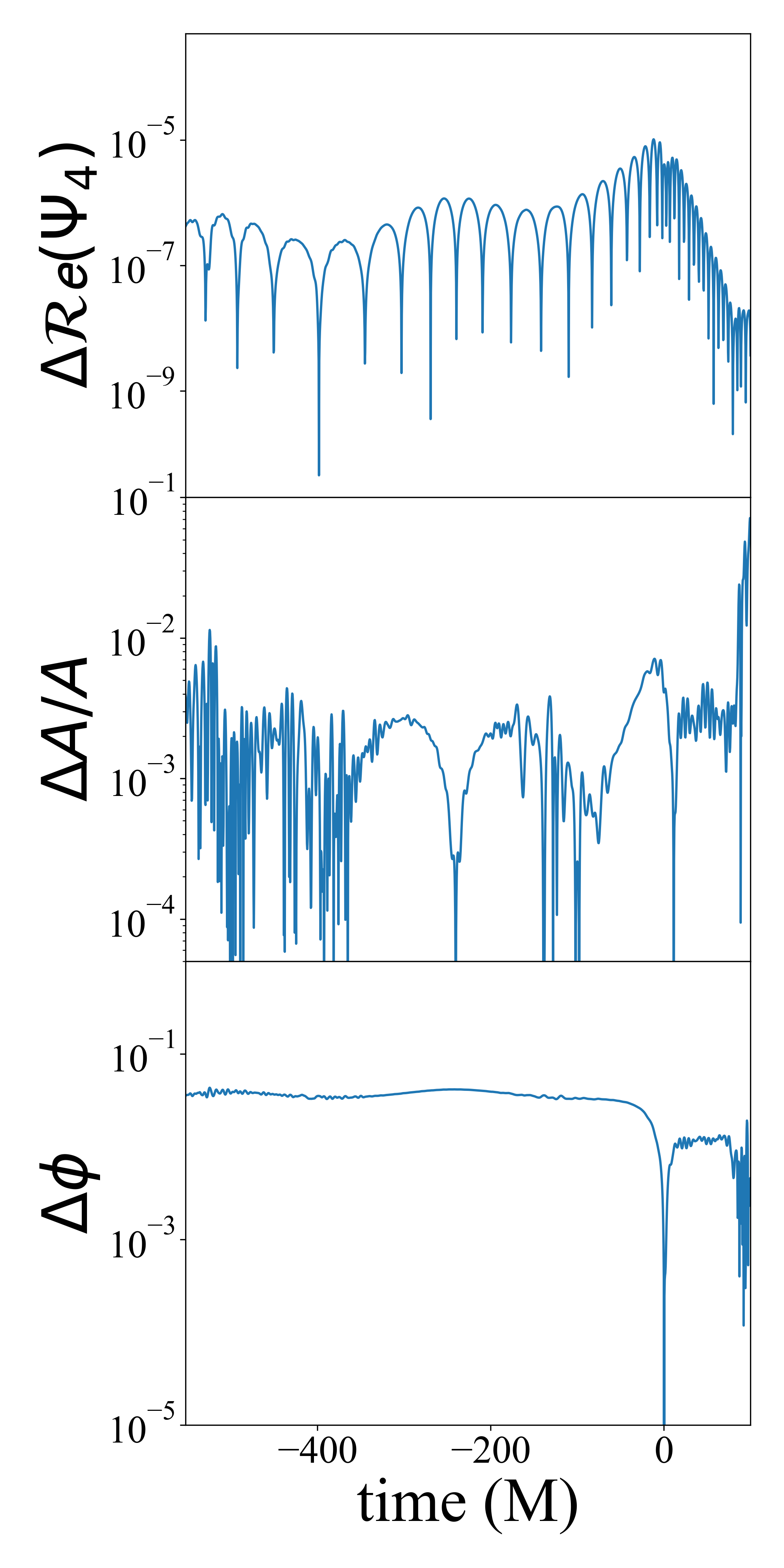}
\caption{}
\label{fig:est_error_eccentric}
\end{subfigure}
\caption{Convergence and error analysis for an equal mass, nonspinning, system with eccentricity of 0.1 and separation 11M.  a) The residuals for the real part, amplitude, and phase of $\Psi_4$ using the high and medium resolutions (blue), the medium and low resolutions (orange), and the high and low resolutions (green). The orange and green lines are scaled according to the convergence factor using Eq. ~\ref{eq:convergence_function}. b) Estimated error in the standard resolution waveform.}
\label{fig:eccentric}
\end{figure*}

We repeat the same methodology for an equal mass, nonspinning, eccentric simulation ($e=0.1$) and show the results in Fig.~\ref{fig:eccentric}.
Figure ~\ref{fig:convergence_eccentric} shows the residuals between the different resolutions with the residuals between the medium and low resolutions and the high and low resolutions being scaled using a convergence factor of $\alpha=2.94$.
Figure ~\ref{fig:est_error_eccentric} shows the estimated error for the standard resolution (48 points across the radius of the \bh{}).
The amplitude error reaches $\sim 0.7\%$  at merger, and the phase error reaches $\sim 0.035$.
Given a mismatch of $\sim 10^{-3}$ between the lowest and highest resolution waveforms, our standard waveform will be indistinguishable from an ``infinite'' resolution, face-on signal up to an \snr{} of $\rho = 161$ and unlikely to cause significant parameter bias up to $\rho = 457$.

\begin{figure*}
\centering
\begin{subfigure}{.45\textwidth}
    \centering
    \includegraphics[width=\linewidth]{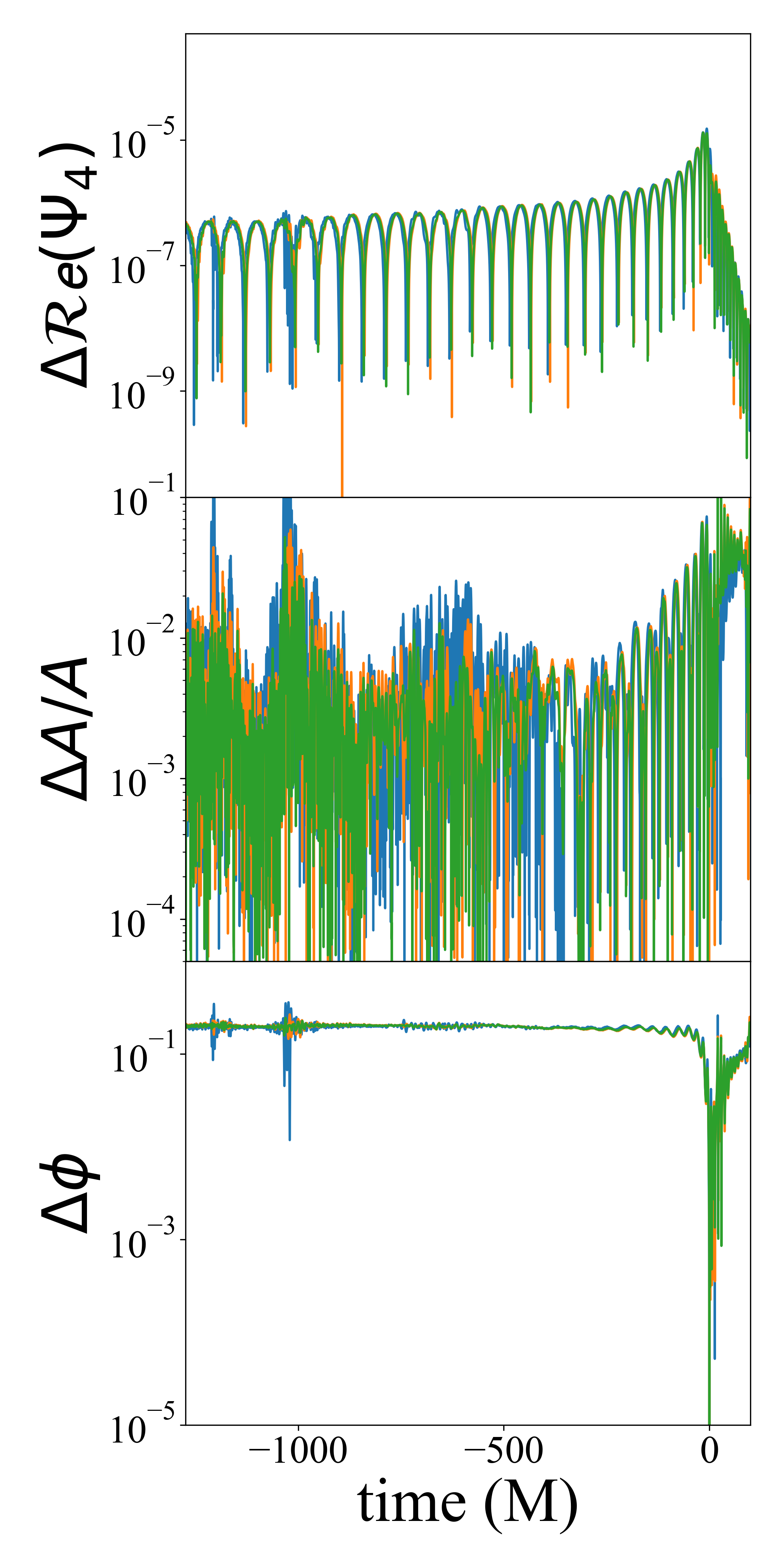}
\caption{}
\label{fig:convergence_q5}
\end{subfigure}
\begin{subfigure}{.45\textwidth}
    \centering
    \includegraphics[width=\linewidth]{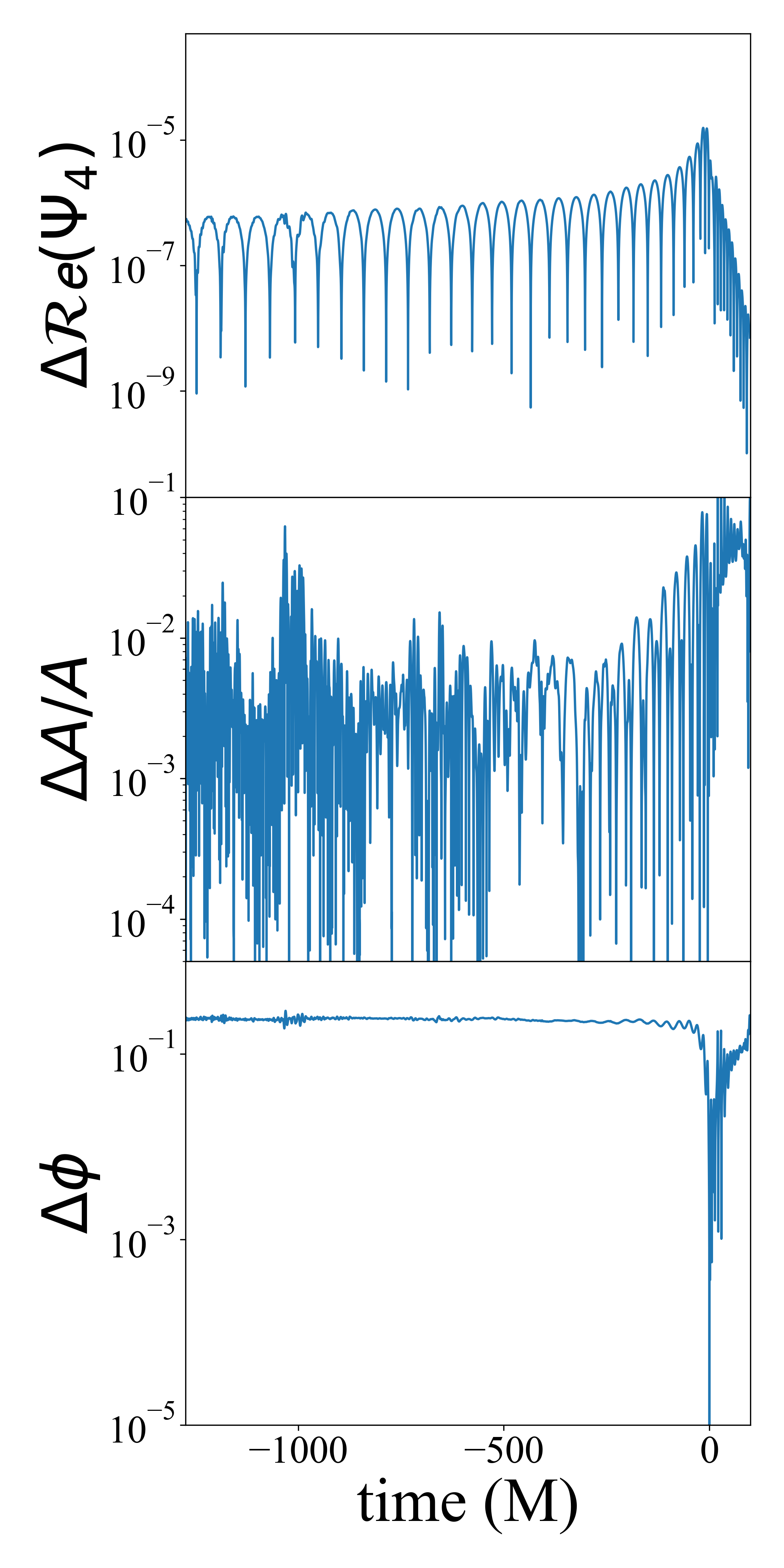}
\caption{}
\label{fig:est_error_q5}
\end{subfigure}
\caption{Convergence and error analysis for a $q=5$, precessing system with separation 11M (MAYA1087).  a) The residuals for the real part, amplitude, and phase of $\Psi_4$ using the high and medium resolutions (blue), the medium and low resolutions (orange), and the high and low resolutions (green). The orange and green lines are scaled according to the convergence factor using Eq. ~\ref{eq:convergence_function}. b) Estimated error in the standard resolution waveform.}
\label{fig:q5}
\end{figure*}

We also perform the same analysis for a precessing, $q=5$ simulation (MAYA1087), shown in Fig.~\ref{fig:q5}, with Figure~\ref{fig:convergence_q5} showing a convergence factor of $\alpha=2.12$ and Figure ~\ref{fig:est_error_q5} showing the estimated error for the standard resolution (48 points across the radius of the \bh{}).
The amplitude error peaks at $\sim 7.5\%$  at merger, and the phase error reaches $\sim 0.22$.
Given a mismatch  of $\sim 10^{-2}$ between the lowest and highest resolution waveforms, our standard waveform will be indistinguishable from an ``infinite'' resolution, face-on signal up to an \snr{} of $\rho = 22.6$ and unlikely to cause significant parameter bias up to $\rho = 63.8$.

A few of the simulations were constructed with courser grids as described in Sec.~\ref{sec:catalog_description}.
In particular,  the suite studying secondary spin described in Sec.~\ref{sec:high_mass_ratio} includes simulations with only 24 points across the finite grid. 
This corresponds to the lowest resolution simulation used in this convergence suite.
A $q=5$ simulation with 24 points across the radius of the finest grid could cause parameter bias above $\rho = 14.7$.

\begin{table*}[t]
\renewcommand\thetable{I} 
\caption{Initial parameters for the 181 new simulations in the MAYA Catalog. The provided spins are
  the dimensionless spin parameters at the beginning of the simulation with the \bh{s} separated
  along the x-axis and the orbital angular momentum in the +z direction. The eccentricity and orbital
  frequency ($M\Omega$) are computed after junk radiation. The cycles are measured from after junk
  radiation until the peak of the sum of the squares of the $\ell=2$ modes. The resolution is defined
  as the number of points per length M on the finest mesh refinement.}
\begin{tabular}{| c | | c | c | c | c | c | c | c | c | c | c | c |}
\hline
Tag & mass ratio & $a_{1x}$ & $a_{1y}$ & $a_{1z}$ & $a_{2x}$ & $a_{2y}$ & $a_{2z}$ & eccentricity & cycles & orbital frequency & resolution (1/M) \\ 
\hline
\hline
MAYA0907 & 1.508 & 0.000 & 0.000 & 0.395 & 0.000 & 0.000 & -0.809 & 0.0199 & 5.4 & 0.0410 & 159.09 \\ 
MAYA0908 & 2.011 & 0.000 & 0.000 & 0.055 & 0.000 & 0.000 & 0.809 & 0.0105 & 8.2 & 0.0376 & 160.00 \\ 
MAYA0909 & 2.000 & 0.000 & 0.000 & 0.200 & 0.000 & 0.000 & 0.000 & 0.0028 & 7.2 & 0.0385 & 159.09 \\ 
MAYA0910 & 1.277 & 0.647 & 0.000 & 0.152 & 0.679 & 0.000 & -0.021 & 0.0059 & 9.6 & 0.0327 & 240.00 \\ 
MAYA0911 & 3.978 & -0.455 & 0.091 & -0.657 & -0.050 & 0.040 & -0.190 & 0.0092 & 6.0 & 0.0344 & 436.36 \\ 
MAYA0912 & 3.438 & 0.000 & 0.000 & 0.192 & 0.000 & 0.000 & 0.693 & 0.0078 & 26.9 & 0.0213 & 140.00 \\ 
MAYA0913 & 1.000 & 0.000 & 0.000 & 0.400 & 0.000 & 0.000 & 0.400 & 0.0118 & 18.0 & 0.0248 & 184.62 \\ 
MAYA0914 & 1.000 & 0.000 & 0.000 & 0.400 & 0.000 & 0.000 & 0.400 & 0.0229 & 17.4 & 0.0251 & 184.62 \\ 
MAYA0915 & 1.000 & 0.000 & 0.000 & 0.400 & 0.000 & 0.000 & 0.400 & 0.0340 & 16.8 & 0.0255 & 184.62 \\ 
MAYA0916 & 1.000 & 0.000 & 0.000 & 0.400 & 0.000 & 0.000 & 0.400 & 0.0449 & 16.1 & 0.0259 & 184.62 \\ 
MAYA0917 & 1.000 & 0.000 & 0.000 & 0.400 & 0.000 & 0.000 & 0.400 & 0.0558 & 15.5 & 0.0262 & 184.62 \\ 
MAYA0918 & 1.000 & 0.000 & 0.000 & 0.400 & 0.000 & 0.000 & 0.400 & 0.0667 & 14.9 & 0.0267 & 184.62 \\ 
MAYA0919 & 1.000 & 0.000 & 0.000 & 0.400 & 0.000 & 0.000 & 0.400 & 0.0777 & 14.3 & 0.0271 & 184.62 \\ 
MAYA0920 & 1.000 & 0.000 & 0.000 & 0.400 & 0.000 & 0.000 & 0.400 & 0.0888 & 13.8 & 0.0275 & 184.62 \\ 
MAYA0921 & 1.000 & 0.000 & 0.000 & 0.400 & 0.000 & 0.000 & 0.400 & 0.0998 & 13.3 & 0.0280 & 184.62 \\ 
MAYA0922 & 1.000 & 0.000 & 0.000 & 0.000 & 0.000 & 0.000 & 0.000 & 0.0126 & 14.4 & 0.0251 & 184.62 \\ 
MAYA0923 & 1.000 & 0.000 & 0.000 & 0.000 & 0.000 & 0.000 & 0.000 & 0.0618 & 11.9 & 0.0267 & 184.62 \\ 
MAYA0924 & 1.000 & 0.000 & 0.000 & 0.000 & 0.000 & 0.000 & 0.000 & 0.0741 & 11.4 & 0.0271 & 184.62 \\ 
MAYA0925 & 1.000 & 0.000 & 0.000 & 0.000 & 0.000 & 0.000 & 0.000 & 0.1106 & 9.7 & 0.0285 & 184.62 \\ 
MAYA0926 & 1.000 & 0.000 & 0.000 & 0.000 & 0.000 & 0.000 & 0.000 & 0.0268 & 13.7 & 0.0253 & 240.00 \\ 
MAYA0927 & 1.000 & 0.000 & 0.000 & 0.000 & 0.000 & 0.000 & 0.000 & 0.0392 & 13.1 & 0.0256 & 240.00 \\ 
MAYA0928 & 1.000 & 0.000 & 0.000 & 0.000 & 0.000 & 0.000 & 0.000 & 0.0514 & 12.4 & 0.0260 & 240.00 \\ 
MAYA0929 & 1.000 & 0.000 & 0.000 & 0.000 & 0.000 & 0.000 & 0.000 & 0.0885 & 10.8 & 0.0272 & 240.00 \\ 
MAYA0930 & 1.000 & 0.000 & 0.000 & 0.000 & 0.000 & 0.000 & 0.000 & 0.1024 & 10.2 & 0.0277 & 240.00 \\ 
MAYA0931 & 1.000 & 0.000 & 0.000 & 0.000 & 0.000 & 0.000 & 0.000 & 0.1356 & 8.6 & 0.0291 & 240.00 \\ 
MAYA0932 & 1.000 & 0.000 & 0.000 & 0.000 & 0.000 & 0.000 & 0.000 & 0.2025 & 6.2 & 0.0277 & 240.00 \\ 
MAYA0933 & 1.000 & 0.000 & 0.000 & 0.000 & 0.000 & 0.000 & 0.000 & 0.2336 & 5.2 & 0.0291 & 240.00 \\ 
MAYA0934 & 1.000 & 0.000 & 0.000 & 0.000 & 0.000 & 0.000 & 0.000 & 0.2909 & 4.1 & 0.0251 & 240.00 \\ 
MAYA0935 & 1.000 & 0.000 & 0.000 & 0.000 & 0.000 & 0.000 & 0.000 & 0.3926 & 1.8 & 0.0222 & 240.00 \\ 
MAYA0936 & 1.000 & 0.000 & 0.000 & 0.000 & 0.000 & 0.000 & 0.000 & 0.6016 & 2.0 & 0.0054 & 240.00 \\ 
MAYA0937 & 2.000 & 0.000 & 0.000 & 0.400 & 0.000 & 0.000 & 0.400 & 0.0034 & 20.5 & 0.0243 & 282.35 \\ 
MAYA0938 & 2.000 & 0.000 & 0.000 & 0.400 & 0.000 & 0.000 & 0.400 & 0.0117 & 19.8 & 0.0246 & 282.35 \\ 
MAYA0939 & 2.000 & 0.000 & 0.000 & 0.400 & 0.000 & 0.000 & 0.400 & 0.0228 & 19.1 & 0.0250 & 282.35 \\ 
MAYA0940 & 2.000 & 0.000 & 0.000 & 0.400 & 0.000 & 0.000 & 0.400 & 0.0339 & 18.4 & 0.0254 & 282.35 \\ 
MAYA0941 & 2.000 & 0.000 & 0.000 & 0.400 & 0.000 & 0.000 & 0.400 & 0.0449 & 17.7 & 0.0257 & 282.35 \\ 
MAYA0942 & 2.000 & 0.000 & 0.000 & 0.400 & 0.000 & 0.000 & 0.400 & 0.0559 & 17.0 & 0.0261 & 282.35 \\ 
MAYA0943 & 2.000 & 0.000 & 0.000 & 0.400 & 0.000 & 0.000 & 0.400 & 0.0670 & 16.4 & 0.0265 & 282.35 \\ 
MAYA0944 & 2.000 & 0.000 & 0.000 & 0.400 & 0.000 & 0.000 & 0.400 & 0.0782 & 15.7 & 0.0269 & 282.35 \\ 
MAYA0945 & 2.000 & 0.000 & 0.000 & 0.400 & 0.000 & 0.000 & 0.400 & 0.0895 & 15.0 & 0.0274 & 282.35 \\ 
MAYA0946 & 2.000 & 0.000 & 0.000 & 0.400 & 0.000 & 0.000 & 0.400 & 0.1008 & 14.4 & 0.0278 & 282.35 \\ 
MAYA0947 & 2.000 & 0.000 & 0.000 & 0.400 & 0.000 & 0.000 & 0.400 & 0.1121 & 13.9 & 0.0283 & 282.35 \\ 
MAYA0948 & 2.000 & 0.000 & 0.000 & 0.000 & 0.000 & 0.000 & 0.000 & 0.0015 & 16.3 & 0.0245 & 240.00 \\ 
MAYA0949 & 2.000 & 0.000 & 0.000 & 0.000 & 0.000 & 0.000 & 0.000 & 0.0135 & 15.6 & 0.0248 & 240.00 \\ 
MAYA0950 & 2.000 & 0.000 & 0.000 & 0.000 & 0.000 & 0.000 & 0.000 & 0.0261 & 14.9 & 0.0252 & 240.00 \\ 
MAYA0951 & 2.000 & 0.000 & 0.000 & 0.000 & 0.000 & 0.000 & 0.000 & 0.0387 & 14.2 & 0.0256 & 240.00 \\ 
MAYA0952 & 2.000 & 0.000 & 0.000 & 0.000 & 0.000 & 0.000 & 0.000 & 0.0512 & 13.5 & 0.0259 & 240.00 \\ 
MAYA0953 & 2.000 & 0.000 & 0.000 & 0.000 & 0.000 & 0.000 & 0.000 & 0.0636 & 12.8 & 0.0263 & 240.00 \\ 
MAYA0954 & 2.000 & 0.000 & 0.000 & 0.000 & 0.000 & 0.000 & 0.000 & 0.0761 & 12.2 & 0.0267 & 240.00 \\ 
MAYA0955 & 2.000 & 0.000 & 0.000 & 0.000 & 0.000 & 0.000 & 0.000 & 0.0887 & 11.6 & 0.0272 & 240.00 \\ 
MAYA0956 & 2.000 & 0.000 & 0.000 & 0.000 & 0.000 & 0.000 & 0.000 & 0.1015 & 11.0 & 0.0276 & 240.00 \\ 
MAYA0957 & 2.000 & 0.000 & 0.000 & 0.000 & 0.000 & 0.000 & 0.000 & 0.1147 & 10.3 & 0.0281 & 240.00 \\ 
MAYA0958 & 2.000 & 0.000 & 0.000 & 0.000 & 0.000 & 0.000 & 0.000 & 0.1233 & 9.6 & 0.0285 & 240.00 \\ 
MAYA0959 & 3.000 & 0.000 & 0.000 & 0.400 & 0.000 & 0.000 & 0.400 & 0.0045 & 23.8 & 0.0240 & 369.23 \\ 
MAYA0960 & 3.000 & 0.000 & 0.000 & 0.400 & 0.000 & 0.000 & 0.400 & 0.0112 & 22.9 & 0.0244 & 369.23 \\ 
MAYA0961 & 3.000 & 0.000 & 0.000 & 0.400 & 0.000 & 0.000 & 0.400 & 0.0225 & 22.0 & 0.0247 & 369.23 \\ 
MAYA0962 & 3.000 & 0.000 & 0.000 & 0.400 & 0.000 & 0.000 & 0.400 & 0.0337 & 21.2 & 0.0250 & 369.23 \\ 
MAYA0963 & 3.000 & 0.000 & 0.000 & 0.400 & 0.000 & 0.000 & 0.400 & 0.0449 & 20.4 & 0.0254 & 369.23 \\ 
\hline
\end{tabular}
\end{table*}

\begin{table*}[t]
\renewcommand\thetable{I} 
\caption{Continuation of table on previous page.}
\begin{tabular}{| c | | c | c | c | c | c | c | c | c | c | c | c |} 
\hline
Tag & mass ratio & $a_{1x}$ & $a_{1y}$ & $a_{1z}$ & $a_{2x}$ & $a_{2y}$ & $a_{2z}$ & eccentricity & cycles & orbital frequency & resolution (1/M) \\ 
\hline
\hline
MAYA0964 & 3.000 & 0.000 & 0.000 & 0.400 & 0.000 & 0.000 & 0.400 & 0.0560 & 19.6 & 0.0258 & 369.23 \\ 
MAYA0965 & 3.000 & 0.000 & 0.000 & 0.400 & 0.000 & 0.000 & 0.400 & 0.0671 & 18.8 & 0.0262 & 369.23 \\ 
MAYA0966 & 3.000 & 0.000 & 0.000 & 0.400 & 0.000 & 0.000 & 0.400 & 0.0783 & 17.9 & 0.0266 & 369.23 \\ 
MAYA0967 & 3.000 & 0.000 & 0.000 & 0.400 & 0.000 & 0.000 & 0.400 & 0.0895 & 17.3 & 0.0270 & 369.23 \\ 
MAYA0968 & 3.000 & 0.000 & 0.000 & 0.400 & 0.000 & 0.000 & 0.400 & 0.1009 & 16.5 & 0.0274 & 369.23 \\ 
MAYA0969 & 3.000 & 0.000 & 0.000 & 0.400 & 0.000 & 0.000 & 0.400 & 0.1123 & 15.8 & 0.0278 & 369.23 \\ 
MAYA0970 & 3.000 & 0.000 & 0.000 & 0.000 & 0.000 & 0.000 & 0.000 & 0.0045 & 18.5 & 0.0244 & 240.00 \\ 
MAYA0971 & 3.000 & 0.000 & 0.000 & 0.000 & 0.000 & 0.000 & 0.000 & 0.0122 & 17.7 & 0.0247 & 240.00 \\ 
MAYA0972 & 3.000 & 0.000 & 0.000 & 0.000 & 0.000 & 0.000 & 0.000 & 0.0250 & 16.9 & 0.0251 & 240.00 \\ 
MAYA0973 & 3.000 & 0.000 & 0.000 & 0.000 & 0.000 & 0.000 & 0.000 & 0.0377 & 16.0 & 0.0255 & 240.00 \\ 
MAYA0974 & 3.000 & 0.000 & 0.000 & 0.000 & 0.000 & 0.000 & 0.000 & 0.0505 & 15.3 & 0.0258 & 240.00 \\ 
MAYA0975 & 3.000 & 0.000 & 0.000 & 0.000 & 0.000 & 0.000 & 0.000 & 0.0632 & 14.5 & 0.0262 & 240.00 \\ 
MAYA0976 & 3.000 & 0.000 & 0.000 & 0.000 & 0.000 & 0.000 & 0.000 & 0.0760 & 13.8 & 0.0267 & 240.00 \\ 
MAYA0977 & 3.000 & 0.000 & 0.000 & 0.000 & 0.000 & 0.000 & 0.000 & 0.0889 & 12.9 & 0.0271 & 240.00 \\ 
MAYA0978 & 3.000 & 0.000 & 0.000 & 0.000 & 0.000 & 0.000 & 0.000 & 0.1020 & 12.2 & 0.0275 & 240.00 \\ 
MAYA0979 & 3.000 & 0.000 & 0.000 & 0.000 & 0.000 & 0.000 & 0.000 & 0.1152 & 11.6 & 0.0280 & 240.00 \\ 
MAYA0980 & 3.000 & 0.000 & 0.000 & 0.000 & 0.000 & 0.000 & 0.000 & 0.1278 & 10.8 & 0.0285 & 240.00 \\ 
MAYA0981 & 4.000 & 0.000 & 0.000 & 0.400 & 0.000 & 0.000 & 0.400 & 0.0032 & 27.1 & 0.0240 & 436.36 \\ 
MAYA0982 & 4.000 & 0.000 & 0.000 & 0.400 & 0.000 & 0.000 & 0.400 & 0.0122 & 26.1 & 0.0243 & 436.36 \\ 
MAYA0983 & 4.000 & 0.000 & 0.000 & 0.400 & 0.000 & 0.000 & 0.400 & 0.0233 & 25.1 & 0.0246 & 436.36 \\ 
MAYA0984 & 4.000 & 0.000 & 0.000 & 0.400 & 0.000 & 0.000 & 0.400 & 0.0345 & 24.1 & 0.0250 & 436.36 \\ 
MAYA0985 & 4.000 & 0.000 & 0.000 & 0.400 & 0.000 & 0.000 & 0.400 & 0.0457 & 23.2 & 0.0254 & 436.36 \\ 
MAYA0986 & 4.000 & 0.000 & 0.000 & 0.400 & 0.000 & 0.000 & 0.400 & 0.0567 & 22.3 & 0.0257 & 436.36 \\ 
MAYA0987 & 4.000 & 0.000 & 0.000 & 0.400 & 0.000 & 0.000 & 0.400 & 0.0678 & 21.3 & 0.0261 & 436.36 \\ 
MAYA0988 & 4.000 & 0.000 & 0.000 & 0.400 & 0.000 & 0.000 & 0.400 & 0.0789 & 20.4 & 0.0265 & 436.36 \\ 
MAYA0989 & 4.000 & 0.000 & 0.000 & 0.400 & 0.000 & 0.000 & 0.400 & 0.0902 & 19.6 & 0.0269 & 436.36 \\ 
MAYA0990 & 4.000 & 0.000 & 0.000 & 0.400 & 0.000 & 0.000 & 0.400 & 0.1017 & 18.6 & 0.0273 & 436.36 \\ 
MAYA0991 & 4.000 & 0.000 & 0.000 & 0.400 & 0.000 & 0.000 & 0.400 & 0.1132 & 17.8 & 0.0278 & 436.36 \\ 
MAYA0992 & 4.000 & 0.000 & 0.000 & 0.000 & 0.000 & 0.000 & 0.000 & 0.0114 & 20.1 & 0.0247 & 400.00 \\ 
MAYA0993 & 4.000 & 0.000 & 0.000 & 0.000 & 0.000 & 0.000 & 0.000 & 0.0264 & 18.9 & 0.0251 & 436.36 \\ 
MAYA0994 & 4.000 & 0.000 & 0.000 & 0.000 & 0.000 & 0.000 & 0.000 & 0.0392 & 18.0 & 0.0255 & 436.36 \\ 
MAYA0995 & 4.000 & 0.000 & 0.000 & 0.000 & 0.000 & 0.000 & 0.000 & 0.0521 & 17.0 & 0.0259 & 436.36 \\ 
MAYA0996 & 4.000 & 0.000 & 0.000 & 0.000 & 0.000 & 0.000 & 0.000 & 0.0650 & 16.1 & 0.0263 & 436.36 \\ 
MAYA0997 & 4.000 & 0.000 & 0.000 & 0.000 & 0.000 & 0.000 & 0.000 & 0.0779 & 15.3 & 0.0267 & 436.36 \\ 
MAYA0998 & 4.000 & 0.000 & 0.000 & 0.000 & 0.000 & 0.000 & 0.000 & 0.0910 & 14.4 & 0.0271 & 436.36 \\ 
MAYA0999 & 4.000 & 0.000 & 0.000 & 0.000 & 0.000 & 0.000 & 0.000 & 0.1044 & 13.4 & 0.0276 & 436.36 \\ 
MAYA1000 & 4.000 & 0.000 & 0.000 & 0.000 & 0.000 & 0.000 & 0.000 & 0.1180 & 12.6 & 0.0280 & 436.36 \\ 
MAYA1001 & 4.000 & 0.000 & 0.000 & 0.000 & 0.000 & 0.000 & 0.000 & 0.1314 & 12.0 & 0.0285 & 436.36 \\ 
MAYA1002 & 15.002 & 0.000 & 0.000 & 0.000 & 0.000 & 0.000 & -0.200 & 0.0046 & 35.4 & 0.0277 & 800.00 \\ 
MAYA1003 & 15.012 & 0.000 & 0.000 & 0.000 & 0.000 & 0.000 & -0.401 & 0.0045 & 35.8 & 0.0277 & 800.00 \\ 
MAYA1004 & 15.039 & 0.000 & 0.000 & 0.000 & 0.000 & 0.000 & -0.603 & 0.0120 & 36.2 & 0.0278 & 800.00 \\ 
MAYA1005 & 15 & 0.000 & 0.000 & 0.000 & 0.000 & 0.000 & -0.8 & - & 27.4 & 0.0327 & 800.00 \\ 
MAYA1006 & 15.002 & 0.000 & 0.000 & 0.000 & 0.000 & 0.000 & 0.200 & 0.0517 & 28.4 & 0.0294 & 800.00 \\ 
MAYA1007 & 15 & 0.000 & 0.000 & 0.000 & 0.000 & 0.000 & 0.4 & - & 30.2 & 0.0275 & 800.00 \\ 
MAYA1008 & 15 & 0.000 & 0.000 & 0.000 & 0.000 & 0.000 & 0.6 & - & 32.6 & 0.0274 & 800.00 \\ 
MAYA1009 & 15.071 & 0.000 & 0.000 & 0.000 & 0.000 & 0.000 & 0.807 & 0.0051 & 41.3 & 0.0278 & 800.00 \\ 
MAYA1010 & 15.000 & 0.000 & 0.000 & 0.000 & 0.000 & 0.000 & 0.000 & 0.0023 & 35.9 & 0.0276 & 800.00 \\ 
MAYA1011 & 1.000 & 0.000 & 0.000 & 0.000 & 0.000 & 0.000 & -0.200 & 0.0037 & 14.0 & 0.0248 & 240.00 \\ 
MAYA1012 & 1.000 & 0.000 & 0.000 & 0.000 & 0.000 & 0.000 & -0.400 & 0.0054 & 13.1 & 0.0250 & 240.00 \\ 
MAYA1013 & 1.001 & 0.000 & 0.000 & 0.000 & 0.000 & 0.000 & -0.601 & 0.0081 & 12.3 & 0.0251 & 240.00 \\ 
MAYA1014 & 1.005 & 0.000 & 0.000 & 0.000 & 0.000 & 0.000 & -0.809 & 0.0106 & 11.4 & 0.0253 & 240.00 \\ 
MAYA1015 & 1.000 & 0.000 & 0.000 & 0.000 & 0.000 & 0.000 & 0.000 & 0.0023 & 14.8 & 0.0247 & 240.00 \\ 
MAYA1016 & 3.000 & 0.000 & 0.000 & 0.000 & 0.000 & 0.000 & -0.200 & 0.0043 & 18.0 & 0.0245 & 240.00 \\ 
MAYA1017 & 3.001 & 0.000 & 0.000 & 0.000 & 0.000 & 0.000 & -0.400 & 0.0033 & 18.0 & 0.0246 & 240.00 \\ 
MAYA1018 & 3.003 & 0.000 & 0.000 & 0.000 & 0.000 & 0.000 & -0.601 & 0.0049 & 18.0 & 0.0246 & 240.00 \\ 
MAYA1019 & 3.018 & 0.000 & 0.000 & 0.000 & 0.000 & 0.000 & -0.809 & 0.0077 & 16.9 & 0.0246 & 240.00 \\ 
MAYA1020 & 3.000 & 0.000 & 0.000 & 0.000 & 0.000 & 0.000 & 0.200 & 0.0045 & 19.0 & 0.0244 & 240.00 \\ 
MAYA1021 & 3.000 & 0.000 & 0.000 & 0.000 & 0.000 & 0.000 & 0.400 & 0.0021 & 20.1 & 0.0243 & 240.00 \\ 
MAYA1022 & 3.003 & 0.000 & 0.000 & 0.000 & 0.000 & 0.000 & 0.601 & 0.0023 & 21.6 & 0.0242 & 240.00 \\ 
MAYA1023 & 3.018 & 0.000 & 0.000 & 0.000 & 0.000 & 0.000 & 0.809 & 0.0046 & 19.9 & 0.0242 & 240.00 \\ 
MAYA1024 & 5.000 & 0.000 & 0.000 & 0.000 & 0.000 & 0.000 & -0.200 & 0.0014 & 24.2 & 0.0243 & 302.40 \\ 
MAYA1025 & 5.001 & 0.000 & 0.000 & 0.000 & 0.000 & 0.000 & -0.400 & 0.0025 & 23.8 & 0.0244 & 302.40 \\ 
\hline
\end{tabular}
\end{table*}

\begin{table*}[t]
\renewcommand\thetable{I} 
\caption{Continuation of table on previous page.}
\begin{tabular}{| c | | c | c | c | c | c | c | c | c | c | c | c |} 
\hline
Tag & mass ratio & $a_{1x}$ & $a_{1y}$ & $a_{1z}$ & $a_{2x}$ & $a_{2y}$ & $a_{2z}$ & eccentricity & cycles & orbital frequency & resolution (1/M) \\ 
\hline
\hline
MAYA1026 & 5.006 & 0.000 & 0.000 & 0.000 & 0.000 & 0.000 & -0.601 & 0.0025 & 23.3 & 0.0244 & 302.40 \\ 
MAYA1027 & 5.036 & 0.000 & 0.000 & 0.000 & 0.000 & 0.000 & -0.811 & 0.0058 & 23.3 & 0.0246 & 302.40 \\ 
MAYA1028 & 5.000 & 0.000 & 0.000 & 0.000 & 0.000 & 0.000 & 0.200 & 0.0027 & 25.1 & 0.0242 & 302.40 \\ 
MAYA1029 & 5.001 & 0.000 & 0.000 & 0.000 & 0.000 & 0.000 & 0.400 & 0.0026 & 25.5 & 0.0242 & 302.40 \\ 
MAYA1030 & 5.006 & 0.000 & 0.000 & 0.000 & 0.000 & 0.000 & 0.601 & 0.0028 & 25.7 & 0.0242 & 302.40 \\ 
MAYA1031 & 5.036 & 0.000 & 0.000 & 0.000 & 0.000 & 0.000 & 0.811 & 0.0130 & 27.5 & 0.0241 & 302.40 \\ 
MAYA1032 & 5.000 & 0.000 & 0.000 & 0.000 & 0.000 & 0.000 & 0.000 & 0.0022 & 24.6 & 0.0243 & 302.40 \\ 
MAYA1033 & 7.000 & 0.000 & 0.000 & 0.000 & 0.000 & 0.000 & -0.200 & 0.0038 & 28.5 & 0.0242 & 360.00 \\ 
MAYA1034 & 7.052 & 0.000 & 0.000 & 0.000 & 0.000 & 0.000 & -0.812 & 0.0112 & 28.5 & 0.0238 & 360.00 \\ 
MAYA1035 & 7.000 & 0.000 & 0.000 & 0.000 & 0.000 & 0.000 & 0.200 & 0.0034 & 32.1 & 0.0242 & 384.00 \\ 
MAYA1036 & 7.002 & 0.000 & 0.000 & 0.000 & 0.000 & 0.000 & 0.400 & 0.0045 & 32.0 & 0.0241 & 384.00 \\ 
MAYA1037 & 7.012 & 0.000 & 0.000 & 0.000 & 0.000 & 0.000 & 0.602 & 0.0054 & 31.3 & 0.0241 & 384.00 \\ 
MAYA1038 & 7.052 & 0.000 & 0.000 & 0.000 & 0.000 & 0.000 & 0.812 & 0.0168 & 37.8 & 0.0240 & 384.00 \\ 
MAYA1039 & 7.000 & 0.000 & 0.000 & 0.000 & 0.000 & 0.000 & 0.000 & 0.0029 & 31.9 & 0.0242 & 384.00 \\ 
MAYA1040 & 4.000 & 0.000 & 0.000 & 0.000 & 0.000 & 0.000 & 0.000 & 0.0018 & 21.1 & 0.0243 & 400.00 \\ 
MAYA1041 & 1.000 & 0.000 & 0.000 & 0.400 & 0.000 & 0.000 & 0.400 & 0.1107 & 12.7 & 0.0285 & 184.62 \\ 
MAYA1042 & 1.000 & 0.400 & 0.000 & 0.000 & 0.400 & 0.000 & 0.000 & 0.0043 & 15.2 & 0.0247 & 184.62 \\ 
MAYA1043 & 1.000 & 0.400 & 0.000 & 0.000 & 0.400 & 0.000 & 0.000 & 0.1208 & 9.4 & 0.0290 & 184.62 \\ 
MAYA1044 & 1.000 & 0.401 & 0.000 & 0.401 & 0.401 & 0.000 & 0.401 & 0.1105 & 12.9 & 0.0284 & 184.62 \\ 
MAYA1045 & 4.546 & 0.080 & 0.266 & -0.543 & 0.466 & -0.093 & -0.039 & 0.1245 & 2.4 & 0.0325 & 480.00 \\ 
MAYA1046 & 4.935 & -0.561 & 0.083 & -0.560 & 0.435 & -0.645 & -0.145 & 0.1219 & 2.3 & 0.0326 & 533.33 \\ 
MAYA1047 & 4.966 & 0.291 & 0.117 & -0.490 & 0.401 & 0.088 & -0.032 & 0.1438 & 3.0 & 0.0318 & 533.33 \\ 
MAYA1048 & 4.994 & -0.407 & 0.284 & -0.604 & -0.027 & -0.010 & -0.713 & 0.1129 & 2.3 & 0.0321 & 533.33 \\ 
MAYA1049 & 4.988 & 0.461 & 0.000 & -0.512 & -0.087 & 0.192 & 0.237 & 0.1600 & 3.2 & 0.0318 & 533.33 \\ 
MAYA1050 & 1.962 & 0.929 & -0.076 & -0.170 & 0.082 & 0.350 & 0.652 & 0.0120 & 11.4 & 0.0314 & 282.35 \\ 
MAYA1051 & 1.259 & 0.222 & 0.246 & -0.040 & -0.313 & -0.209 & 0.728 & 0.0044 & 11.6 & 0.0311 & 208.70 \\ 
MAYA1052 & 2.376 & -0.757 & -0.501 & -0.129 & -0.506 & -0.560 & 0.179 & 0.0060 & 11.7 & 0.0307 & 320.00 \\ 
MAYA1053 & 4.840 & 0.358 & 0.576 & -0.318 & 0.485 & 0.041 & 0.209 & 0.0028 & 10.5 & 0.0307 & 533.33 \\ 
MAYA1054 & 5.313 & 0.117 & 0.454 & -0.424 & 0.264 & -0.268 & 0.341 & 0.0022 & 10.5 & 0.0309 & 600.00 \\ 
MAYA1055 & 2.271 & 0.696 & -0.062 & -0.554 & -0.015 & -0.530 & 0.127 & 0.0057 & 8.1 & 0.0315 & 300.00 \\ 
MAYA1056 & 1.205 & 0.047 & -0.316 & 0.301 & -0.918 & 0.102 & -0.066 & 0.0078 & 10.9 & 0.0310 & 200.00 \\ 
MAYA1057 & 1.084 & -0.163 & -0.139 & -0.246 & -0.046 & -0.201 & -0.745 & 0.0023 & 11.2 & 0.0252 & 192.00 \\ 
MAYA1058 & 1.163 & 0.019 & -0.146 & -0.563 & -0.033 & 0.017 & -0.789 & 0.3935 & 9.9 & 0.0256 & 200.00 \\ 
MAYA1059 & 1.164 & -0.006 & -0.147 & -0.563 & -0.029 & 0.025 & -0.795 & 0.0032 & 9.9 & 0.0255 & 200.00 \\ 
MAYA1060 & 1.901 & 0.722 & 0.000 & 0.321 & -0.009 & 0.015 & -0.308 & 0.0019 & 18.0 & 0.0246 & 266.67 \\ 
MAYA1061 & 1.926 & -0.640 & -0.206 & 0.170 & -0.129 & 0.228 & -0.247 & 0.0032 & 17.0 & 0.0247 & 266.67 \\ 
MAYA1062 & 2.714 & -0.443 & -0.407 & 0.556 & -0.300 & 0.585 & -0.311 & 0.0051 & 22.6 & 0.0243 & 342.86 \\ 
MAYA1063 & 1.989 & 0.010 & 0.009 & 0.809 & -0.037 & -0.031 & 0.205 & 0.0036 & 23.0 & 0.0242 & 282.35 \\ 
MAYA1064 & 2.001 & 0.203 & 0.612 & -0.489 & 0.767 & 0.000 & -0.257 & 0.0035 & 12.8 & 0.0251 & 282.35 \\ 
MAYA1065 & 1.991 & 0.232 & -0.740 & 0.173 & -0.211 & 0.435 & 0.058 & 0.0039 & 17.8 & 0.0246 & 282.35 \\ 
MAYA1066 & 1.989 & 0.772 & 0.222 & 0.092 & -0.184 & -0.175 & 0.005 & 0.0030 & 17.1 & 0.0246 & 282.35 \\ 
MAYA1067 & 2.998 & -0.454 & 0.602 & -0.293 & 0.160 & 0.000 & 0.768 & 0.0044 & 18.4 & 0.0246 & 369.23 \\ 
MAYA1068 & 2.992 & 0.345 & -0.361 & -0.633 & -0.303 & -0.192 & -0.632 & 0.0056 & 12.0 & 0.0251 & 369.23 \\ 
MAYA1069 & 2.983 & -0.176 & 0.331 & -0.717 & 0.084 & 0.012 & -0.207 & 0.0033 & 12.1 & 0.0251 & 369.23 \\ 
MAYA1070 & 2.987 & -0.435 & 0.609 & -0.265 & 0.499 & 0.000 & -0.054 & 0.0045 & 17.3 & 0.0246 & 369.23 \\ 
MAYA1071 & 2.983 & -0.511 & -0.496 & -0.383 & 0.012 & -0.010 & 0.010 & 0.0038 & 15.7 & 0.0248 & 369.23 \\ 
MAYA1072 & 2.983 & -0.515 & -0.549 & -0.297 & 0.000 & 0.004 & -0.016 & 0.0040 & 16.8 & 0.0247 & 369.23 \\ 
MAYA1073 & 2.984 & -0.572 & -0.194 & -0.538 & -0.365 & 0.026 & -0.065 & 0.0042 & 14.4 & 0.0250 & 369.23 \\ 
MAYA1074 & 3.001 & -0.592 & -0.105 & -0.542 & -0.706 & -0.291 & -0.261 & 0.0069 & 13.4 & 0.0252 & 369.23 \\ 
MAYA1075 & 3.001 & -0.661 & -0.142 & -0.425 & -0.768 & -0.206 & -0.017 & 0.0066 & 15.6 & 0.0250 & 369.23 \\ 
MAYA1076 & 3.002 & -0.708 & -0.054 & -0.362 & -0.698 & -0.141 & 0.373 & 0.0045 & 16.6 & 0.0248 & 369.23 \\ 
MAYA1077 & 2.984 & -0.746 & 0.040 & -0.310 & 0.129 & -0.046 & 0.411 & 0.0038 & 17.6 & 0.0247 & 369.23 \\ 
MAYA1078 & 2.983 & -0.750 & 0.100 & -0.286 & 0.001 & 0.000 & -0.229 & 0.0042 & 15.5 & 0.0249 & 369.23 \\ 
MAYA1079 & 3.002 & -0.758 & 0.053 & -0.277 & -0.616 & 0.496 & 0.174 & 0.0052 & 16.6 & 0.0247 & 369.23 \\ 
MAYA1080 & 3.002 & 0.064 & 0.765 & -0.257 & 0.296 & 0.713 & 0.243 & 0.0030 & 17.2 & 0.0245 & 369.23 \\ 
MAYA1081 & 3.002 & 0.116 & 0.649 & -0.467 & 0.628 & -0.014 & -0.511 & 0.0028 & 13.9 & 0.0248 & 369.23 \\ 
MAYA1082 & 3.002 & 0.264 & -0.437 & -0.628 & 0.098 & 0.717 & -0.364 & 0.0040 & 12.0 & 0.0251 & 369.23 \\ 
MAYA1083 & 3.005 & 0.294 & 0.547 & -0.463 & 0.554 & 0.589 & 0.046 & 0.0064 & 14.8 & 0.0249 & 369.23 \\ 
MAYA1084 & 4.042 & -0.901 & 0.080 & 0.047 & -0.223 & -0.004 & 0.699 & 0.0053 & 24.8 & 0.0244 & 480.00 \\ 
MAYA1085 & 4.396 & -0.376 & -0.127 & -0.702 & -0.011 & -0.006 & -0.047 & 0.0046 & 13.3 & 0.0252 & 480.00 \\ 
MAYA1086 & 4.944 & -0.426 & 0.055 & -0.596 & 0.052 & -0.025 & -0.373 & 0.0073 & 15.1 & 0.0252 & 533.33 \\ 
MAYA1087 & 4.983 & -0.365 & -0.055 & -0.644 & -0.001 & 0.001 & -0.084 & 0.0060 & 15.0 & 0.0251 & 533.33 \\ 
\hline
\end{tabular}
\end{table*}

\clearpage

\bibliography{ms}
\end{document}